\pdfoutput=1  % tells arXiv to build with pdflatex
\documentclass[sigconf,nonacm]{acmart}

\usepackage{tabularx}
\usepackage{multirow}
\usepackage{tikz}
\usetikzlibrary{arrows.meta,positioning,calc,fit,backgrounds,patterns}
\usepackage{pgfplots}
\pgfplotsset{compat=1.18}
\usepgfplotslibrary{groupplots}

\newcommand{\clsinner}[1]{{\ttfamily\def\_{\textunderscore\allowbreak}#1}}
\newcommand{\cls}[1]{\relax\ifmmode\text{\clsinner{#1}}\else\clsinner{#1}\fi}
\newcommand{\M}[1]{\textsf{M#1}}    % label observability channels M1, M2, M3
\newcommand{\fiveel}{\emph{Five Elements Inc.}}

\tikzset{
  box/.style={draw, rounded corners=1pt, align=center, font=\scriptsize,
              inner sep=3pt, line width=0.5pt},
  flow/.style={-{Stealth[length=4pt]}, line width=0.5pt},
  note/.style={font=\tiny\itshape, text=black!60, align=center},
}

\begin{document}

% ================================================================= FRONT ==
\title{Inference-Layer Security: Defending Against Adversarial Inference
       and Infrastructure Abuse}

\author{Keifer Lee}
\email{keifer@nymlr.com}
\affiliation{%
  \institution{New York Machine Learning Research Guild (NYMLR)}
  \city{New York}
  \state{NY}
  \country{USA}}

\begin{abstract}
Operating a large language model (LLM) as a service requires more than inference
infrastructure: the provider must also defend against adversarial interactions that seek
to exploit the service, including jailbreaking for harmful use, sophisticated denial of
service, and distillation attacks. We study this problem at the inference layer, using a
hypothetical frontier lab, \fiveel, as a running example. Because no public labelled
dataset of adversarial LLM usage exists, we introduce a structural causal model (SCM)
that generates a realistically grounded, labelled dataset of user-sessions, with
coordinated multi-account campaigns, platform feedback, and three tiers of label
observability. On this dataset we train a practical gradient-boosted detector that
classifies each user-session as benign or malicious and, if malicious, by attack type.
Against oracle labels the detector very nearly solves the binary task (AUPRC $0.993$),
yet against the operational labels a real Trust \& Safety team would hold, the same model
scores an AUPRC of only $0.313$: the detector is more accurate than the labels used to
evaluate it. For attack-type attribution, a naive argmax is dominated by the $98\%$ benign
prior (macro-F1 $0.295$), whereas a simple thresholded decision engine raises macro-F1 to
$0.489$ without sacrificing accuracy. The dataset is publicly released.
\end{abstract}


\maketitle

% =========================================================== INTRODUCTION ==
\section{The Need to Defend AI Systems}
\label{sec:intro}

A growing share of white-collar workflows now depends on AI in one form or another:
roughly $40\%$ of Coinbase's daily code is reportedly AI-generated~\cite{mexc2025coinbase},
Cloudflare lets AI agents register their own cloud accounts, subscriptions, and domains
end-to-end~\cite{infoworld2026cloudflare}, and Anthropic's Claude has been integrated
directly into Excel~\cite{anthropicnews}. Like any engineered system, these deployments
carry exploitable vulnerabilities, yet adoption is outpacing security: GenAI usage in
organisations rose from $33\%$ in 2023 to $71\%$ in 2024~\cite{cloudflareaisec} without a
commensurate security stack, including in services where consistency, security, and
stability are non-negotiable. Coinbase's CEO has stated that ``non-technical teams are now
shipping production code''~\cite{pymnts2026vibe} at the exchange, days before its trading
engine went down~\cite{deepdive2026coinbase}; Cloudflare reports that ``85\% of IT decision
makers report that employees are adopting AI tools faster than their IT teams can assess
them''~\cite{cloudflareaisec}. This haste persists despite METR's randomised controlled
trial, which found that experienced developers were $19\%$ \emph{slower} with AI tooling
while believing they were $20\%$ faster~\cite{becker2025metr}.

Conventional software is already fraught with exploits, known or otherwise. Google's
Threat Intelligence Group tracked 90 zero-days exploited in the wild in 2025 alone, a
$15\%$ increase year-over-year~\cite{bleeping2026zerodays}, and these are only the ones
we know about; the population of \emph{living} zero-days---unknown to the vendor,
exploitable for an extended period, sometimes never patched at all~\cite{wikizeroday}---is
by definition uncountable. Software is a formal system layered on other formal systems, and
it inherits their incompleteness in the form of edge cases nobody thought to check. The
result is a steady stream of high-profile incidents, such as the MOVEit/GoAnywhere/Cleo
managed-file-transfer cascade~\cite{acroniszeroday} and CVE-2026-32202, a zero-click
Windows Shell vulnerability actively exploited by APT28 via weaponised LNK
files~\cite{helpnet2026lpe}. Deep generative models expand this attack surface further,
introducing vectors that are largely unknown to the public that relies on these services
and that are harder to detect and guard against. Notable attack vectors from recent months
include:

\begin{enumerate}
  \item \textbf{Poison 250 documents in a corpus of billions to backdoor any LLM.} A joint
    Anthropic / UK AI Security Institute / Alan Turing Institute study found that as few
    as 250 adversarial documents---${\sim}0.00016\%$ of a 13B model's training
    tokens---reliably install a backdoor, regardless of model
    scale~\cite{anthropic2025poison,turing2025poison}.
  \item \textbf{Trivial jailbreak via past-tense framing and persona injection.} The
    OpenAI--Anthropic joint safety evaluation found Claude models were ``most vulnerable
    to the past tense jailbreak''~\cite{openai2025jointeval}, and the Mexican government
    attacker simply told Claude it was working a bug bounty programme to unlock the full
    kill chain~\cite{hawkeye2026mexico}.
  \item \textbf{Distillation attacks at industrial scale.} Anthropic disclosed that
    DeepSeek, Moonshot and MiniMax ran ${\sim}24{,}000$ fraudulent accounts and 16M+
    exchanges against Claude to extract chain-of-thought training
    data~\cite{anthropic2026distill}---MiniMax pivoted to a new Claude model within 24
    hours of its release~\cite{ailearned2026distill}.
  \item \textbf{Agentic abuse / weaponisation as a multiplier.} Anthropic's report on
    disrupting AI espionage describes a Chinese state-sponsored actor (GTG-1002) achieving
    $80$--$90\%$ autonomy of a cyber-espionage campaign with only 4--6 human touchpoints;
    their model was making ``thousands of requests, often multiple per
    second''~\cite{anthropic2025espionage}.
  \item \textbf{AI-authored zero-days in the wild.} Google's GTIG published a case where
    they could fingerprint a 2FA-bypass zero-day exploit as AI-generated from its
    hallucinated CVSS score and textbook docstrings~\cite{helpnet2026aizeroday}.
\end{enumerate}

\noindent Recent high-profile attacks include:

\begin{enumerate}
  \item \textbf{The Mexican government breach (Dec 2025 -- Feb 2026).} A single operator
    used Claude Code + GPT-4.1 to compromise 10 government bodies and exfiltrate
    ${\sim}150$\,GB / 195 million identities---Claude executed ${\sim}75\%$ of the remote
    commands~\cite{securityweek2026mexico}.
  \item \textbf{Linux kernel LPE + Chrome zero-day chain (early 2026).} A high-severity
    Linux kernel local-privilege-escalation~\cite{helpnet2026lpe} landed alongside four
    actively exploited Chrome zero-days in the first quarter of 2026
    alone~\cite{securityaffairs2026dawn}---Google was the second-most-targeted vendor of
    2025.
  \item \textbf{North Korean IT-worker fraud, AI-augmented.} Anthropic disclosed that DPRK
    operatives were using Claude to fraudulently secure and maintain remote employment at
    Fortune 500 tech companies~\cite{anthropic2025misuse}, funnelling salaries back to the
    regime's weapons programmes.
  \item \textbf{LLMjacking at scale---Operation Bizarre Bazaar.} Pillar Security's
    honeypots recorded ${\sim}972$ attack sessions per day over 40 days targeting exposed
    LLM endpoints~\cite{pillar2026bizarre}, with a 9-minute credential-to-first-exploitation
    time, sold via the silver.inc marketplace on Telegram/Discord~\cite{securityweek2026bizarre}.
  \item \textbf{Supply-chain compromise of LiteLLM and Trivy.} In March 2026, the TeamPCP /
    UNC6780 group embedded the SANDCLOCK credential stealer into LiteLLM and Trivy
    builds~\cite{helpnet2026aizeroday}, exfiltrating AWS and GitHub secrets---and, by
    extension, the AI environments those keys unlocked.
\end{enumerate}

These examples show that attacks are varied and have been demonstrated to great effect.
How, then, can model providers prevent their services from being abused? No single
measure suffices; defence requires a suite of measures at each stage of the model
lifecycle (Figure~\ref{fig:serving}). In this paper we focus on the serving side, for
three reasons:

\begin{enumerate}
  \item Model weaknesses are a function of many factors, including the training data,
    training procedure, infrastructure, and architecture (for example, energy-based models
    are less prone to hallucination than autoregressive LLMs, while weaker language models
    lack the coherence needed for fine-grained control). All of these are fixed at
    training time; once a model is deployed, its priors cannot be changed without a costly
    retraining run.
  \item The serving side is far easier to control: the serving platform has full
    visibility and well-tested deterministic levers, such as API rate limiting,
    harmful-content filtering, IP/device blocklisting, and velocity throttling. The same
    toolbox that payment networks have used for decades to keep card fraud at
    ${\sim}7$\,bps of GMV largely transfers.
  \item Serving-side interventions generalise across model types and attack types. A rate
    limit is indifferent to whether the model behind it is GPT-5.5 or a fine-tuned
    Qwen~3.6, and an attacker's clustered account graph has the same shape regardless.
\end{enumerate}

% ----------------------------------------------------------- FIGURE 1 ----
\begin{figure}[t]
\centering
\scriptsize
\setlength{\tabcolsep}{3pt}
\renewcommand{\arraystretch}{1.15}
\begin{tabularx}{\columnwidth}{@{}>{\raggedright\arraybackslash}p{0.23\columnwidth}
                                   >{\raggedright\arraybackslash}X
                                   >{\raggedright\arraybackslash}p{0.30\columnwidth}@{}}
\toprule
\textbf{Stage} & \textbf{Attacks at this stage} & \textbf{Defensive levers} \\
\midrule
\multicolumn{3}{@{}l}{\tiny\itshape baked-in: no runtime lever} \\
\textbf{Training data} \newline {\tiny corpus curation}
  & data poisoning (250 docs); sock puppets; PII injection
  & \emph{none---corpus is frozen} \\
\textbf{Training / RLHF} \newline {\tiny pre-train + tune}
  & backdoor via fine-tune; alignment failures; RLHF gaming
  & \emph{none---weights are frozen} \\
\midrule
\multicolumn{3}{@{}l}{\tiny\itshape structural runtime (deploy / infra) --- scope of this paper begins here} \\
\textbf{Deployment} \newline {\tiny supply chain}
  & LiteLLM/Trivy compromise; SANDCLOCK credential stealer
  & SBOM; image signing; reproducible builds \\
\textbf{Serving infra} \newline {\tiny gateway, isolation}
  & cross-tenant KV leak; credential theft; LLMjacking
  & tenant isolation; key rotation; TLS pinning; gateway rate limits \\
\midrule
\multicolumn{3}{@{}l}{\tiny\itshape live runtime (per-request / session)} \\
\textbf{Request layer} \newline {\tiny prompts, filters}
  & prompt injection; past-tense jailbreak; probes
  & pre-filter: perplexity; post-filter: safety score \\
\textbf{User session} \newline {\tiny behaviour, ladder}
  & distillation queries (16M); DoS; agentic abuse
  & behavioural detection; intervention ladder \\
\bottomrule
\end{tabularx}
\caption{The serving stack and where exploits enter at each stage: the model can be
poisoned via its training corpus; the infrastructure can be compromised for jailbreaks or
data leaks (e.g.\ leaking system prompts, or other users' prompts); and the user level can
be exploited to serve adversarial results, jailbreak, or distill the model.}
\label{fig:serving}
\Description{A table-style diagram of six serving stages, from training data to user session, listing the attacks that fire at each stage and the defensive levers available.}
\end{figure}

Throughout, we take the perspective of a hypothetical frontier LLM provider, \fiveel,
and ask how it can serve its models responsibly while defending against abuse by
adversaries. This paper makes the following contributions:
\begin{itemize}
  \item a taxonomy, the \emph{Ladder of Abstraction}, for building an inference-layer
    defence stack rung by rung, from raw event streams to multi-agent games
    (Section~\ref{sec:setup});
  \item a structural causal model (SCM) that simulates coordinated adversarial campaigns
    against a serving platform, with realistic label delay, account-level label noise,
    and three tiers of label observability (Sections~\ref{sec:scm}--\ref{sec:labels}); the
    resulting dataset is public~\cite{daoistdurian2026dataset};
  \item a practical GBDT detector with a tempered class-weighting recipe, and a family of
    thresholded decision engines for attack-type attribution under extreme class
    imbalance (Sections~\ref{sec:training}--\ref{sec:results}).
\end{itemize}

% ============================================================ TASK SETUP ==
\section{Defending Five Elements Inc.}
\label{sec:setup}

\begin{figure}[t]
\centering
\begin{tikzpicture}[
  n/.style={box, minimum width=2.45cm, minimum height=0.95cm, text width=2.3cm},
  x=1cm, y=1cm]
  \node[n] (users) at (0,0)     {\textbf{1 $\cdot$ Users}\\[-1pt]{\tiny benign + malicious}};
  \node[n] (app)   at (2.8,0)   {\textbf{2 $\cdot$ Service app}\\[-1pt]{\tiny IP + velocity checks}};
  \node[n] (serve) at (5.6,0)   {\textbf{3 $\cdot$ Serving layer}\\[-1pt]{\tiny gen + pre/post filter}};
  \node[n] (resp)  at (5.6,-1.5){\textbf{4 $\cdot$ Response or block}\\[-1pt]{\tiny return, flag, or suspend}};
  \node[n] (back)  at (2.8,-1.5){\textbf{back to user}\\[-1pt]{\tiny served or rejected}};
  \node[n] (db)    at (5.6,-3.0){\textbf{Transaction DB}\\[-1pt]{\tiny user $\cdot$ session $\cdot$ txn}};
  \node[n] (etl)   at (2.8,-3.0){\textbf{5 $\cdot$ ETL pipeline}\\[-1pt]{\tiny clean $\to$ warehouse}};
  \node[box, minimum width=2.45cm, text width=2.3cm, minimum height=2.45cm, line width=0.9pt]
       (ladder) at (0,-2.25)
       {\textbf{Intervention ladder}\\[2pt]
        {\tiny 1 $\cdot$ CAPTCHA\\ 2 $\cdot$ rate limit\\ 3 $\cdot$ enh.\ monitoring\\
               4 $\cdot$ suspend\\ 5 $\cdot$ revoke key}};
  \node[box, fill=black!8, minimum width=8.05cm, text width=7.8cm, minimum height=0.95cm,
        line width=0.9pt] (ts) at (2.8,-4.6)
       {\textbf{6 $\cdot$ Trust \& Safety}\\[-1pt]
        {\tiny review analytics $\cdot$ confirm labels $\cdot$ issue interventions $\cdot$ update policy}};

  \draw[flow] (users) -- (app);
  \draw[flow] (app) -- (serve);
  \draw[flow] (serve) -- (resp);
  \draw[flow] (resp) -- (back);
  \draw[flow] (resp) -- (db);
  \draw[flow] (db) -- (etl);
  \draw[flow] (etl) -- (etl |- ts.north);
  % fast loop
  \draw[flow, line width=1.1pt] (ladder |- ts.north) -- (ladder.south);
  \draw[flow, line width=1.1pt] (ladder.north) -- (users.south);
  % slow loop
  \draw[flow, dashed] (ts.east) -- ++(0.35,0) |- (serve.east);

  % legend
  \draw[line width=1.1pt] (-1.2,-5.45) -- ++(0.5,0)
     node[right, font=\tiny] {fast loop: interventions per account (ms--daily)};
  \draw[dashed] (-1.2,-5.75) -- ++(0.5,0)
     node[right, font=\tiny] {slow loop: tune thresholds, reweight features, retrain (weekly--quarterly)};
\end{tikzpicture}
\caption{The complete \fiveel\ serving pipeline, with its intervention (fast) and
policy (slow) feedback loops.}
\label{fig:tsloops}
\Description{Box-and-arrow diagram of the request path from users through the service app, serving layer and response, into the transaction database and ETL pipeline, feeding a Trust and Safety team that drives a fast intervention loop back to users and a slow policy loop back to the serving layer.}
\end{figure}
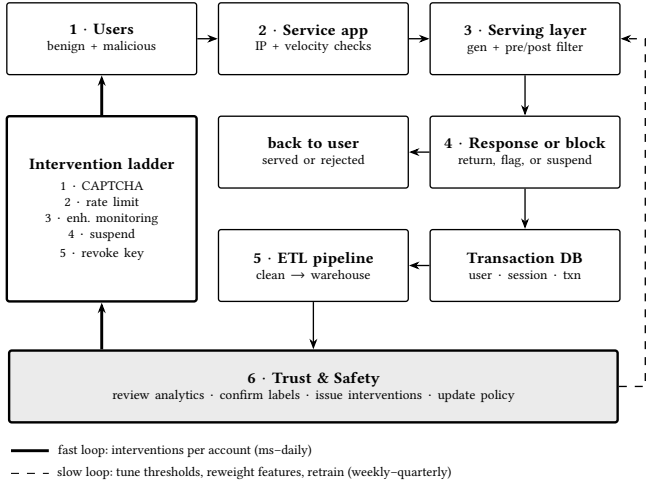

Figure~\ref{fig:tsloops} summarises our task setup at \fiveel\ Platform Trust \& Safety
(T\&S), a standard architecture that provides context for the problem statement. The
goal is to \emph{give legitimate users a low-friction experience while identifying and
blocking adversarial actors}.

% ----------------------------------------------------------- FIGURE 2 ----

Achieving this goal raises several questions:
\begin{enumerate}
  \item What interventions should the platform apply, and how should they be enacted?
  \item How can adversarial actors be identified reliably when each \emph{type} presents a
    different profile and behaviour (e.g.\ a bot farm vs.\ a jailbreak)? Is a binary
    benign/malicious classification sufficient?
  \item How should benign and malicious be defined, given that, for example, a genuine
    researcher can resemble a distillation attacker?
  \item At what \emph{level} should the platform intervene: user, session, or transaction?
\end{enumerate}

\begin{figure}[t]
\centering
\begin{tikzpicture}[x=1cm, y=1cm]
  \foreach \lvl/\name/\desc [count=\i from 0] in {
      L0/Raw event streams/{transactions, sessions, infra / net metadata},
      L1/Signals \& features/{IP / device velocity, diurnal patterns},
      L2/Point predictors/{GBDT / XGBoost, tabular nets},
      L3/Uncertainty estimation/{GPs, DBN / HMM beliefs, Bayes nets},
      L4/Memoryless decisioning/{multi-arm bandits, $\varepsilon$-greedy, UCB},
      L5/{Sequential / memory decisioning}/{MDP / RL, contextual bandits},
      L6/{Adversarial games, partial obs.}/{Stackelberg, POMDP},
      L7/Noisy multi-agent games/{many strategic, imperfect actors at once}} {
    \pgfmathsetmacro{\y}{0.74*\i}
    \ifnum\i=2
      \draw[fill=black!10, line width=1.1pt, rounded corners=1pt] (0,\y) rectangle (6.2,\y+0.62);
      \node[anchor=west, font=\scriptsize\bfseries] at (0.12,\y+0.42) {\lvl\ $\cdot$ \name\ \normalfont$\leftarrow$ \itshape this paper};
    \else
      \draw[line width=0.5pt, rounded corners=1pt] (0,\y) rectangle (6.2,\y+0.62);
      \node[anchor=west, font=\scriptsize\bfseries] at (0.12,\y+0.42) {\lvl\ $\cdot$ \name};
    \fi
    \node[anchor=west, font=\tiny, text=black!65] at (0.12,\y+0.16) {\desc};
  }
  \draw[-{Stealth[length=5pt]}, line width=0.6pt, black!60] (-0.35,0) -- (-0.35,5.8);
  \node[rotate=90, font=\tiny\itshape, text=black!60] at (-0.6,2.9) {each rung depends on the one below};
  \node[font=\tiny\itshape, text=black!60, anchor=west] at (6.3,5.5) {strategic};
  \node[font=\tiny\itshape, text=black!60, anchor=west] at (6.3,0.3) {raw};
  \draw[dotted, black!50] (6.62,0.55) -- (6.62,5.3);
\end{tikzpicture}
\caption{The Ladder of Abstraction: L0 raw event streams $\to$ L7 noisy multi-agent
games. Build each rung before climbing.}
\label{fig:ladder}
\Description{Eight stacked rungs from L0 raw event streams to L7 noisy multi-agent games, with L2 point predictors highlighted as the starting point.}
\end{figure}
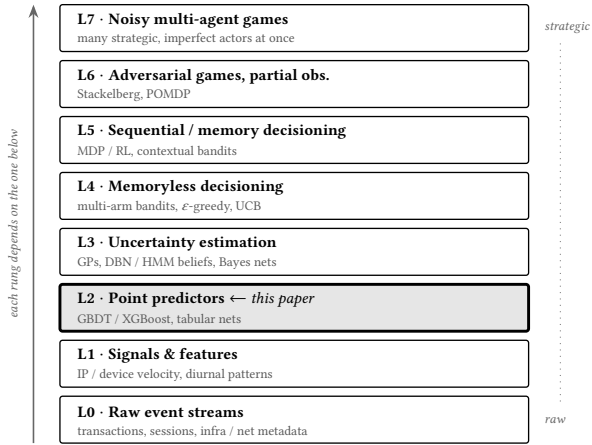

To structure these questions we use the \textbf{Ladder of Abstraction}
(Figure~\ref{fig:ladder}), a solution roadmap that ranges from the most fundamental unit,
raw event streams at L0, to noisy, partially observable, multi-agent games at L7.

% ----------------------------------------------------------- FIGURE 3 ----

In this paper we address L2, point-predictor models such as
XGBoost~\cite{chen2016xgboost}. Reliable point predictions are a prerequisite for the
higher rungs, since each rung consumes the one below it: a bandit (L4) needs a reward
signal, and that reward is only as trustworthy as the point estimate (L2) and the
uncertainty around it (L3) that feed it. Likewise, a Stackelberg
solver~\cite{wikistackelberg} (L6) reasoning about an adversary's best response needs a
belief state (L3) to reason over. Decisions built on poor estimates are confidently wrong,
which is the most damaging failure mode in an adversarial setting.

\subsection{Problem Formulation}

Concretely, our task is as follows: given a user-session, determine whether it is benign
($0$) or adversarial ($1$), and if it is malicious, what kind of bad actor it is:
\begin{equation}
  y_{\text{adv}} \in \{0,1\}, \qquad y_{\text{type}} \in \mathcal{C},
\end{equation}
\begin{equation}
\mathcal{C} = \left\{\begin{array}{@{}l@{\quad}l@{}}
  0{:}\ \cls{benign}        & 5{:}\ \cls{dos} \\
  1{:}\ \cls{distillation}  & 6{:}\ \cls{credential\_abuse} \\
  2{:}\ \cls{jailbreak}     & 7{:}\ \cls{data\_extraction} \\
  3{:}\ \cls{bot\_farm}     & 8{:}\ \cls{agentic\_misuse} \\
  4{:}\ \cls{harmful\_use}  &
\end{array}\right\},
\end{equation}
where each non-benign class maps to a documented real-world surface
(Table~\ref{tab:classes}).

% ------------------------------------------------------------ TABLE 1 ----
\begin{table*}[t]
\caption{The attack-type taxonomy $\mathcal{C}$: what each class is, which team at
\fiveel\ a detection is routed to, and what a miss actually costs.}
\label{tab:classes}
\small
\begin{tabularx}{\textwidth}{@{}l X l X@{}}
\toprule
\textbf{Class} & \textbf{What it is} & \textbf{Routed to} & \textbf{What a miss actually costs} \\
\midrule
\cls{distillation}      & Capability extraction by competitors & Legal \& Policy & a competitor ships your exfiltrated capability \\
\cls{data\_extraction}  & Training-data / PII / system-prompt exfiltration & Legal \& Policy & training-data, PII, or system-prompt leakage \\
\cls{jailbreak}         & Safety bypass (past-tense framing, persona injection) & Trust \& Safety & a safety bypass loose in the wild \\
\cls{harmful\_use}      & Extortion, malware, ransomware-as-a-service & Trust \& Safety & real-world harm---extortion, malware, RaaS \\
\cls{agentic\_misuse}   & Tool-chain abuse as an attack platform & Trust \& Safety & your own tools turned into an attack platform \\
\cls{bot\_farm}         & Coordinated fake-account infrastructure & Platform Integrity & a fake-account fleet scaling unchecked \\
\cls{dos}               & Unbounded consumption / token flooding & Platform Integrity & unbounded consumption, token flooding \\
\cls{credential\_abuse} & Stolen-key / reverse-proxy LLMjacking & Security & stolen-key / reverse-proxy LLMjacking \\
\cls{multi\_surface}$^{\dagger}$ & Coordinated multi-vector campaign & Threat Intelligence & a coordinated multi-vector campaign \\
\cls{benign}            & Honest usage & No action & \emph{(the flip side: a good user slapped with needless friction)} \\
\bottomrule
\multicolumn{4}{@{}l}{\footnotesize $^{\dagger}$ A campaign-level (group) property; out of scope for the per-session predictor.}
\end{tabularx}
\end{table*}

\paragraph{Why aggregate to the user-session?}
A single transaction is almost information-free in isolation: one prompt that trips the
safety filter could be a curious researcher, a typo, or the first probe of a jailbreak.
The discriminating signal lives in the \emph{shape} of a session---the trajectory of
perplexity, the accumulation of refusals, the diurnal rhythm, the inter-request timing
variance (low for bots, high for humans). The user-session preserves these behavioural
shapes while remaining granular enough for targeted interventions. A coarser unit
(whole-account, all-time) loses the onset of an attack; a finer one (per-transaction)
amounts to classifying noise.

The dataset also contains a tenth type at the \emph{campaign level},
\cls{multi\_surface}: a coordinated operation spanning several of the above types. Because
it is a \emph{group} property requiring cross-account aggregation, it is out of scope for a
per-session point predictor, and we defer it to the higher rungs.

Formally, given a feature vector $\mathbf{x} \in \mathbb{R}^d$ for a session, we estimate
\begin{equation}
  \hat{y}_{\text{adv}} = f_{\text{adv}}(\mathbf{x}) \in [0,1], \qquad
  \hat{\mathbf{y}}_{\text{type}} = f_{\text{mc}}(\mathbf{x}) \in \Delta^{8},
\end{equation}
where $\Delta^{8}$ is the 8-simplex over the 9 classes (probabilities summing to 1), and
$f_{\text{adv}}$ collapses to the complement of the benign-class probability in the
simplest single-model framing:
\begin{equation}
  \hat{y}_{\text{adv}} = 1 - \big[f_{\text{mc}}(\mathbf{x})\big]_{\cls{benign}}.
\end{equation}

\subsection{Metrics}

For the binary task we report precision--recall curves and, for threshold-agnostic
comparison, the area under them. For attack-type attribution, a multi-class task, we
report per-class precision and recall and their macro and micro aggregates. For the binary
case, with $\mathrm{TP}$, $\mathrm{FP}$, $\mathrm{FN}$ the true positives, false positives
and false negatives:
\begin{equation}
\text{precision} = \frac{\mathrm{TP}}{\mathrm{TP}+\mathrm{FP}}, \qquad
\text{recall} = \frac{\mathrm{TP}}{\mathrm{TP}+\mathrm{FN}},
\end{equation}
\begin{equation}
\text{AUPRC} = \int_{0}^{1} \text{precision}(\text{recall})\; d(\text{recall}).
\end{equation}
For the multi-class case we compute these per class $k$ in a one-vs-rest fashion and
then aggregate over $K$ classes. With $\mathrm{TP}_k, \mathrm{FP}_k, \mathrm{FN}_k$ the
counts for class $k$, $\text{precision}_k$ and $\text{recall}_k$ follow as above, and
\begin{align}
  \text{macro-P} &= \frac{1}{K}\sum_{k=1}^{K} \text{precision}_k
      && \text{(every class counts equally)}, \\
  \text{micro-P} &= \frac{\sum_{k} \mathrm{TP}_k}{\sum_{k} (\mathrm{TP}_k + \mathrm{FP}_k)}
      && \text{(every sample counts equally)}.
\end{align}
Macro treats \cls{agentic\_misuse} (rare) as importantly as \cls{benign} (the majority),
whereas micro lets the majority class dominate.

\subsection{Data Requirements}

Existing LLM-serving datasets, such as BurstGPT~\cite{wang2024burstgpt} and
ServeGen~\cite{xiang2026servegen}, capture serving patterns and transactions of generic
interactions but do not meet the requirements of this task, which also needs:
\begin{enumerate}
  \item \textbf{Labels} indicating whether each transaction or user-session is benign or
    adversarial and, if adversarial, of what \emph{type}. To our knowledge no public
    labelled dataset of this kind exists, and providers are strongly disincentivised from
    releasing one: publishing confirmed-adversarial traffic reveals how much got through
    (reputational damage), creates legal and regulatory exposure, \emph{and} hands
    adversaries a labelled map of the behavioural signatures that detectors catch.
  \item \textbf{Scale}, on the order of millions of transactions at least, because
    adversarial interactions are rare relative to benign ones. At a deliberately
    conservative transaction-level rate of $p = 0.01\%$ (1 in $10{,}000$), each malicious
    example comes with ${\sim}10{,}000$ benign ones, so collecting $1{,}000$ malicious
    training rows requires scanning
    \begin{equation}
      N_{\text{total}} = \frac{1{,}000}{p} = \frac{1{,}000}{10^{-4}} = 10^{7}
      \ \text{transactions}.
    \end{equation}
\end{enumerate}
The choice of $0.01\%$ is conservative. Open-web figures are far higher: Arkose Labs found
that $73\%$ of web/app traffic in 2023 was bot and fraud-farm
traffic~\cite{securityweek2023bots}, and Statista puts 2024 bad-bot traffic at around
$37\%$~\cite{statista2024bots}. These, however, are \emph{traffic}-level figures on the
open internet. At the \emph{account} level on a friction-gated platform (payment, KYC,
regional restrictions), the rate should be far lower, since these controls filter out
most automated and low-effort abuse; we assume on the order of a few percent of
accounts at most. Allowing for the fact that active accounts generate more activity, we
assume the transaction-level rate for a given platform lies in the $0.01\%$--$5\%$ band,
and adopt its lower end. Given these requirements and the absence of a suitable open
dataset, we construct a synthetic one.

% ================================================================== SCM ==
\section{Synthetic Data Generation}
\label{sec:scm}

The simplest way to synthesise a dataset is to choose a set of features, define a
distribution for each, and sample each column independently. This ignores how features
correlate and covary. A more faithful generator conditions features on one another,
maintains consistent session-to-session behaviour for each user over time, and simulates
cross-account coordination in multi-account campaigns.

For example, a user-session associated with a bot farm likely has an outlying velocity
footprint, correlated intra-cluster activity cadence, potentially a high degree of device
or IP sharing with other bot instances from the same source, and so on. Each of these
correlations only makes sense in the context of the \emph{other} sessions in the same
cluster---sampling them independently is a category error. Concretely, the value of
feature $j$ on session $s$ for user $u$ at time $t$ is not a free-standing random
variable; it is conditioned on three axes at once:
\begin{equation}
x^{(j)}_{u,s,t} = f_j\Big(
  \underbrace{\mathrm{pa}_j(x_{u,s,t})}_{\text{\tiny intra-session}},\,
  \underbrace{\mathbf{h}_{u,<t}}_{\text{\tiny trajectory}},\,
  \underbrace{\mathbf{c}_{C(u),t}}_{\text{\tiny campaign}}\Big)
  + \varepsilon^{(j)}_{u,s,t},
\label{eq:feature}
\end{equation}
where $\mathrm{pa}_j(\cdot)$ is the set of causal parents of feature $j$ \emph{within} the
same session (e.g.\ \cls{total\_tokens} depends on \cls{n\_requests}), $\mathbf{h}_{u,<t}$
is a sufficient statistic over user $u$'s prior sessions (escalation, coverage saturation,
baseline geo, etc.), $\mathbf{c}_{C(u),t}$ is the shared state of the campaign $C(u)$ that
$u$ belongs to (e.g.\ jailbreak success, coordinated burst tick, domain distillation
progress), and $\varepsilon^{(j)}$ is irreducible per-session noise. Each axis is
necessary: without the first, features are uncorrelated; without the second, accounts
have no memory; without the third, a ``bot farm'' is just $N$ independent accounts with no
shared signature.

We implement this generator as a structural causal model
(SCM)~\cite{wikicausalmodel,pearl2009overview}.\footnote{The modern SCM formalism is largely due to Judea
Pearl~\cite{wikipearl}, crystallised in \emph{Causality: Models, Reasoning, and
Inference}~\cite{pearl2009causality} and the do-calculus that earned him the 2011 Turing
Award. The underlying idea---distinguishing structural equations from regression
equations---traces back further to Sewall Wright's path analysis (1921) and the Cowles
Commission econometricians of the 1940s.}

Rather than sampling each column independently, an SCM represents features as nodes in a
directed acyclic graph, with an edge from each cause to its effect, and generates each
feature as a function of its parents plus exogenous noise. Sampling in topological order
yields synthetic instances whose features co-vary as real ones do: a bot farm's session is
not a set of independent draws but the downstream consequence of \emph{being a bot farm}
(low cadence variance, shared infrastructure, and so on). Because the structure is causal
rather than merely correlational, the model also supports \emph{interventions}: clamping a
node to a value, severing its incoming edges, and asking, for example, ``what would this
session look like if it were not rate-limited?''. Such counterfactual
queries~\cite{huszar2019counterfactuals} are impossible with independent sampling and
enable counterfactual analysis in future work. Our SCM is organised in three concentric
layers, mirroring the three conditioning axes of Eq.~\eqref{eq:feature}. A full treatment
of the SCM is beyond the scope of this paper; below we give an overview sufficient to
interpret the dataset.

\subsection{Case Study: A Coordinated Distillation Campaign}

We illustrate the generative process end-to-end on a single distillation campaign, in
which a coordinated group of actors (or a single operator controlling many automated
agents) attempts to extract high-value data from the model. The exact mechanistic
components of the SCM are abstracted, since they are one of many possible instantiations
of the workflow described here. We further restrict attention to \emph{epistemic}
distillation, in which the attacker systematically tiles the model's knowledge across a
domain space, accumulating coverage slot by slot. A structurally distinct variant,
\emph{orchestration} distillation, feeds the model elaborate multi-tool agentic tasks and
captures the resulting action chains for imitation learning; it has a very different
signature (deep tool chains, near-zero query entropy, potentially higher tool-use rate,
and a preference for strongly agentic models).

The generative process has two components: (1) \emph{definition and instantiation}, which
sets the parameters of the simulated world and its actors; and (2) the \emph{rollout},
which produces the tick-by-tick user-sessions of each account in the campaign.

% ----------------------------------------------------------- FIGURE 4 ----
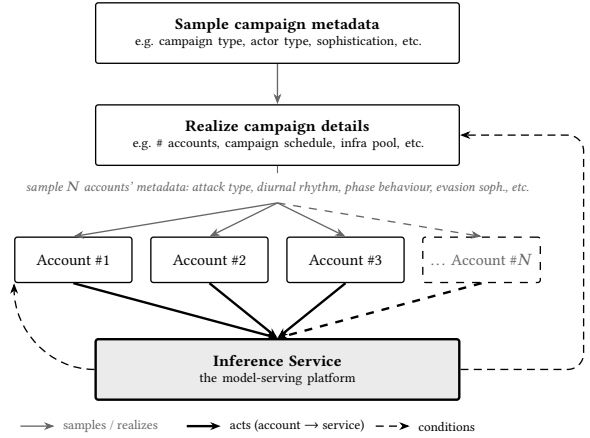
\begin{figure}[t]
\centering
\begin{tikzpicture}[x=1cm, y=1cm,
  acct/.style={box, minimum width=1.55cm, minimum height=0.6cm}]
  \node[box, text width=4.6cm, minimum height=0.8cm] (meta) at (0,0)
    {\textbf{Sample campaign metadata}\\[-1pt]{\tiny e.g.\ campaign type, actor type, sophistication, etc.}};
  \node[box, text width=4.6cm, minimum height=0.8cm] (det) at (0,-1.35)
    {\textbf{Realize campaign details}\\[-1pt]{\tiny e.g.\ \# accounts, campaign schedule, infra pool, etc.}};
  \node[note] (fan) at (0,-2.05) {sample $N$ accounts' metadata: attack type, diurnal rhythm, phase behaviour, evasion soph., etc.};
  \node[acct] (a1) at (-2.7,-3.0) {Account \#1};
  \node[acct] (a2) at (-0.9,-3.0) {Account \#2};
  \node[acct] (a3) at ( 0.9,-3.0) {Account \#3};
  \node[acct, dashed, text=black!60] (aN) at ( 2.7,-3.0) {\dots\ Account \#$N$};
  \node[box, fill=black!8, line width=0.9pt, text width=4.6cm, minimum height=0.8cm] (svc) at (0,-4.45)
    {\textbf{Inference Service}\\[-1pt]{\tiny the model-serving platform}};

  \draw[flow, black!60] (meta) -- (det);
  \draw[black!60] (det.south) -- (fan.north);
  \foreach \a in {a1,a2,a3} \draw[flow, black!60] (fan.south) -- (\a.north);
  \draw[flow, black!60, dashed] (fan.south) -- (aN.north);
  \foreach \a in {a1,a2,a3} \draw[flow, line width=0.9pt] (\a.south) -- (svc.north);
  \draw[flow, line width=0.9pt, dashed] (aN.south) -- (svc.north);
  % conditioning feedback
  \draw[flow, densely dashed] (svc.west) to[out=180, in=270] (a1.south west);
  \draw[flow, densely dashed, rounded corners=6pt] (svc.east) -- (4.05,-4.45) -- (4.05,-1.35) -- (det.east);

  \draw[flow, black!60] (-3.4,-5.2) -- ++(0.45,0) node[right, font=\tiny] {samples / realizes};
  \draw[flow, line width=0.9pt] (-1.2,-5.2) -- ++(0.45,0) node[right, font=\tiny] {acts (account $\to$ service)};
  \draw[flow, densely dashed] (1.3,-5.2) -- ++(0.45,0) node[right, font=\tiny] {conditions};
\end{tikzpicture}
\caption{The SCM's generative pipeline. Campaign metadata is sampled, then realized into
concrete campaign details, which fan out into $N$ account realizations. Each account acts
on the inference service; the service conditions both the accounts and the campaign in
return. The ``etc.''\ marks a deliberately simplified slice---the full SCM samples far more
per box.}
\label{fig:scmdag}
\Description{Campaign metadata is sampled and realized into campaign details, which fan out into N accounts; the accounts act on an inference service, which conditions the accounts and campaign in return.}
\end{figure}

\subsubsection{Defining and instantiating the campaign universe}

Before any account interacts with the model, the simulator sets up the world in which it
will act. This step lays the groundwork for all three axes of Eq.~\eqref{eq:feature} (the
intra-session parents $\mathrm{pa}_j$, the account trajectory $\mathbf{h}_{u,<t}$, and the
shared campaign context $\mathbf{c}_{C(u),t}$) and is, at its core, a four-layer directed
acyclic graph (Figure~\ref{fig:scmdag}).

At the top are two root draws, the campaign type $C$ and the actor type $A$, which fix the
strategic context: \emph{what} kind of attack is simulated and \emph{who} runs it. From
these we draw a \emph{sophistication score} $s$, a scalar that governs nearly every
downstream behavioural parameter. $C$, $A$ and $s$ are then realised into the
\emph{campaign details}, the once-per-campaign quantities such as fleet size, phase
schedule and infrastructure pool, which in turn fan out into \emph{per-account
realisations}, the leaf draws unique to each account. Once instantiated, an account
\emph{acts} on \fiveel's serving stack, and the platform acts back: every rate limit,
CAPTCHA, or monitoring flag conditions the account's future behaviour, and sophisticated
campaigns can take evasive action to circumvent the platform's interventions. This
act/condition loop is the only part of the process that is not a one-shot sample; we
describe it tick by tick in the rollout below.

The description that follows is not exhaustive. The parameter values reflect our
instantiation of the scenario, set from a combination of domain priors and published
estimates, and should be read as illustrative rather than definitive.

\paragraph{Campaign type.} $C \sim \operatorname{Cat}(\boldsymbol{\theta})$ draws which
attack surface this campaign pursues. Distillation is the modal case at $30\%$
(Table~\ref{tab:campaignprior}), reflecting a world where systematic knowledge extraction
is a more common commercial motive than outright harmful use or disruption.

\begin{table}[t]
\caption{Campaign-type prior $\boldsymbol{\theta}$.}
\label{tab:campaignprior}
\small
\begin{tabular}{@{}lr@{\qquad}lr@{}}
\toprule
\cls{distillation}     & 0.30 & \cls{data\_extraction} & 0.10 \\
\cls{harmful\_use}     & 0.20 & \cls{dos}              & 0.07 \\
\cls{jailbreak}        & 0.15 & \cls{multi\_surface}   & 0.05 \\
\cls{credential\_abuse}& 0.10 & \cls{bot\_farm}        & 0.03 \\
\bottomrule
\end{tabular}
\end{table}

\paragraph{Actor type.} $A \sim \operatorname{Cat}(\boldsymbol{\phi})$ identifies who is
running the campaign, with $\boldsymbol{\phi} = (0.30, 0.25, 0.25, 0.10, 0.10)$ for
(\cls{criminal\_org}, \cls{competitor}, \cls{script\_kiddie}, \cls{nation\_state},
\cls{insider}). Criminal organisations are the modal adversary; nation-states are rare but
disproportionately capable.

\paragraph{Sophistication.} $s \mid A, C$ is sampled last and is the most consequential of
the three---it encodes how capable and disciplined this particular operator is:
\begin{equation}
s \mid A, C \sim \operatorname{Beta}\big(\alpha_{A,C}, \beta_{A,C}\big)\Big|_{[s_{\min},\, s_{\max}]}.
\end{equation}
The Beta prior varies by actor type, reflecting our beliefs about each adversary class's
capability floor and ceiling. For distillation campaigns:
\begin{equation}
s \mid A \sim \begin{cases}
\operatorname{Beta}(8,2)\ \ [\mathbb{E}[s]=0.80] & \cls{nation\_state} \\
\operatorname{Beta}(5,3)\ \ [\mathbb{E}[s]=0.63] & \cls{competitor} \\
\operatorname{Beta}(4,4)\ \ [\mathbb{E}[s]=0.50] & \cls{criminal\_org} \\
\operatorname{Beta}(3,3)\ \ [\mathbb{E}[s]=0.50] & \cls{insider} \\
\operatorname{Beta}(1,5)\ \ [\mathbb{E}[s]=0.17] & \cls{script\_kiddie}
\end{cases}
\end{equation}
For distillation campaigns specifically, $s$ is then clamped to $[0.40, 0.95]$---even a
script kiddie running a distillation operation needs to be minimally organised, and nobody
operates with perfect omniscience. Sophistication propagates to nearly every downstream
parameter, including infrastructure reuse, account-creation spread, TLS spoofing,
evasion-response complexity, load-redistribution capability, and the replacement speed of
burnt accounts.

Given these campaign-level parameters, we sample and realise a single distillation
campaign; again, we present only a subset of its key states.

\paragraph{Number of accounts.}
\begin{equation}
n = \max\big(1,\ \operatorname{NegBin}(r{=}3, p{=}0.3) + 5\big), \qquad \mathbb{E}[n] \approx 12.
\end{equation}
The negative binomial (not Poisson) is chosen for its over-dispersion: most operations
deploy a modest cohort, but the heavy right tail allows occasional large fleets. Here $r$
is the stopping parameter and $p$ the per-trial success probability; the $+5$ offset
enforces a floor of genuine coordination. This is an intrinsic property of the
distillation type itself---distillation must tile a large domain space, so it is
inherently multi-account---and is independent of $A$ and $s$.

\paragraph{Phase schedule.} A sampled, accumulated timeline from $t_{\text{start}}$:
\begin{equation}
\begin{aligned}
\tau_{\texttt{setup}}    &\sim U(1,5), &
\tau_{\texttt{warmup}}   &\sim \operatorname{Exp}(\lambda^{-1}{=}7), \\
\tau_{\texttt{probing}}  &\sim U(3,10), &
\tau_{\texttt{cooldown}} &\sim U(1,5), \\
\tau_{\texttt{active}}   &\sim \operatorname{LogNormal}(3.5, 0.5) &&(\text{median} \approx 33\,\text{d}).
\end{aligned}
\end{equation}
Durations are sampled per phase and accumulated into absolute end-days. Crucially, the
setup and warmup phases deliberately emit \emph{benign cover traffic}---a distillation
account looks like an ordinary casual user for its entire early life, and only reveals its
template-driven, systematic behaviour from probing onward (Figure~\ref{fig:phase}).

\paragraph{Termination.} The campaign ends in \emph{success} once aggregate domain
coverage reaches $0.80$, or \emph{aborts} once the suspended fraction of its fleet reaches
$0.70$.

\paragraph{Account-creation spread.}
\begin{equation}
t_{\text{create}} \sim \mathcal{N}\big(t_{\text{start}},\ \sigma = (1-s)\cdot 14\big),
\qquad t_{\text{start}} \sim U(0,\ 0.85\,H).
\end{equation}
Accounts are spawned clustered around the campaign's start day $t_{\text{start}}$ (drawn
over the first $85\%$ of the horizon $H$ so it has room to run). The spread narrows with
sophistication: $s \to 1$ provisions the whole fleet within a tight window for a tighter
campaign cadence---deliberately, often well ahead of activation during the benign
setup/warmup cover phase---while $s \to 0$ scatters registrations across ${\sim}2$ weeks.
This creates a genuine tension: a tight burst \emph{raises}
registration velocity (many sibling registrations in one window), so the sophisticated
operator pairs the burst with low infra reuse (near-unique IPs per account) and hides it
inside the cover-traffic window before any adversarial behaviour invites scrutiny.

\paragraph{Infra pool size and reuse.}
\begin{equation}
n_{\text{IP}} \sim \operatorname{Poi}(5),\ \
n_{\text{dev}} \sim \operatorname{Poi}(8),\ \
n_{\text{email}} \sim \operatorname{Poi}(10);\ \
\text{reuse} = 1 - s.
\end{equation}
The campaign provisions a shared pool of IPs, device fingerprints, and email domains. The
reuse factor $1-s$ governs how aggressively accounts share that pool: high $s$ $\to$ low
reuse $\to$ near-unique infra per account $\to$ a faint shared-IP / shared-device
footprint (hard to catch); low $s$ $\to$ heavy reuse $\to$ a dense, easily clustered
footprint. Sophistication ($s > 0.6$) additionally unlocks TLS-fingerprint spoofing.

\paragraph{Load redistribution.}
\begin{equation}
\text{mult} = \min\Big(3.0,\ \frac{1}{1-f_{\text{burn}}}\Big), \qquad \text{applied only if } s > 0.5.
\label{eq:mult}
\end{equation}
When accounts are suspended (burn fraction $f_{\text{burn}}$), only sophisticated
campaigns ($s > 0.5$) compensate by redistributing the lost throughput across surviving
siblings, scaling each survivor's activity by up to $3\times$. Unsophisticated campaigns
simply absorb the loss and slow down.

\paragraph{Account realisations.} From the campaign definition, individual accounts are
spawned to populate its roster.\footnote{The \cls{multi\_surface} campaign is the most
sophisticated in the taxonomy: its roster is a deliberate mixture of roles---e.g.\ jailbreaker
($35\%$), distillation ($30\%$), agentic misuse ($20\%$), and cover traffic ($15\%$)---and
it is restricted to the nation-state sophistication tier ($s \in [0.70, 1.0]$). The
variety is intentional: it simulates the realistic operational security of a well-funded
adversary who cycles attack surfaces to frustrate pattern-matching defences.} Each then
draws its own realized values from the latent distributions laid out above---its personal
sophistication, how many sessions it runs per day and how many requests within each
session, the average inter-request time, its prompt perplexity and query-structure
entropy, the epistemic domain coverage it targets both per session and in coordination
with its siblings, its behaviour in each phase, its evasion-response complexity, its
diurnal rhythm, and further latent variables.

\paragraph{Label arrival.} Label arrival for each user-session is drawn from a log-normal
delay distribution, to mimic realistic label discovery and annotation at \fiveel:
\begin{equation}
\Pr(\text{ever labeled}) = 0.10, \quad
\text{delay} \sim \operatorname{LogNormal}(\mu{=}4.5, \sigma{=}0.5),
\end{equation}
i.e.\ a median delay of $e^{4.5} \approx 90$ days for distillation. Distillation
sessions are hard to flag: they overlap substantially with benign user types such as the
diligent autodidact, and no single indicator is decisive, so reliable identification
requires accumulating and analysing historical sessions over many days. Even then, in our
setup only $10\%$ of such sessions are ever identified, with a median delay of 90 days.

Table~\ref{tab:overlap} illustrates the overlap by comparing an active distillation
session with a benign power user: on more than half of the listed features, the two
profiles are difficult to distinguish.

% ------------------------------------------------------------ TABLE 3 ----
\begin{table*}[t]
\caption{Configuration of an active distillation session versus a benign power user.}
\label{tab:overlap}
\small
\begin{tabular}{@{}llll@{}}
\toprule
\textbf{Feature} & \textbf{Distillation (active)} & \textbf{Benign power user} & \textbf{Separation} \\
\midrule
\cls{sessions\_per\_day}             & ${\sim}4$/day                & ${\sim}5$/day               & overlap \\
\cls{n\_requests\_per\_session}      & ${\sim}33$/session, heavy    & ${\sim}25$/session, heavy   & overlap \\
\cls{is\_api}                        & $95\%$ API                   & $25\%$ API                  & \textbf{separable} \\
\cls{avg\_inter\_request\_time\_sec} & ${\sim}3$\,s, regular        & ${\sim}12$\,s, varied       & partial \\
\cls{std\_inter\_request\_time\_sec} & ${\sim}8$\,s jitter, robotic & ${\sim}45$\,s jitter, bursty & \textbf{key signal} \\
\cls{avg\_prompt\_perplexity}        & ${\sim}0.0$, tight           & ${\sim}0.2$, looser         & overlap \\
\cls{std\_prompt\_perplexity}        & ${\sim}0.15$, template-driven & ${\sim}0.8$, varied phrasing & \textbf{separable} \\
\cls{query\_structure\_entropy}      & ${\sim}0.25$, narrow/repetitive & ${\sim}0.43$, varied     & \textbf{separable} \\
\cls{avg\_safety\_score}             & ${\sim}0.03$, benign-looking & ${\sim}0.04$, benign-looking & overlap \\
\cls{domain\_coverage\_delta}        & ${\sim}0.50$, aggressive new coverage & ${\sim}0.20$, narrow focus & \textbf{separable} \\
\cls{total\_tokens}                  & ${\sim}1.8$k/session, high   & ${\sim}3.0$k/session, high  & overlap \\
\bottomrule
\end{tabular}
\end{table*}

The table captures a single moment: $C = \cls{distillation}$, $\phi_t = \cls{active}$, an
account mid-campaign. In practice these distributions shift over time: the same account
behaves very differently on the first day of warmup than forty days into active
extraction, and differently again when rate-limited. Formally, for each account $u$ on day
$t$:
\begin{equation}
P(\mathbf{x}_{u,t} \mid \mathcal{Z}) = \prod_{j \in \mathrm{topo}(\mathcal{G})}
  P\Big(x^{(j)}_{u,t} \,\Big|\, \mathrm{pa}_j,\ \theta^{(j)}(\mathcal{Z})\Big),
\end{equation}
\begin{equation}
\mathcal{Z} = \big(C,\ \phi_t,\ \mathbf{h}_{u,<t},\ \mathbf{c}_{C(u),t},\ \mathrm{Interv}_{u,t}\big).
\end{equation}
Here $\mathcal{Z}$ is a context bundle assembled each day from the campaign configuration,
the account's accumulated history, the campaign's shared coordination state, and the
interventions currently applied to the account. The per-node parameters $\theta^{(j)}$
start from a base that depends on $(C, \phi_t)$ \emph{alone}, which is then passed through
a three-stage modifier stack, each stage incorporating one further component of
$\mathcal{Z}$ (written after the semicolon):
\begin{multline}
\theta^{(j)}(\mathcal{Z}) = \mathcal{T}_{\text{traj}}\Big(
  \mathcal{T}_{\text{camp}}\big(
  \mathcal{T}_{\text{interv}}\big(\theta^{(j)}_{\text{base}}(C,\phi_t);\ \mathrm{Interv}_{u,t}\big)
  ; \\
  \mathbf{c}_{C(u),t}\big);\ \mathbf{h}_{u,<t}\Big).
\end{multline}
The base carries $(C, \phi_t)$, and the three stages incorporate the remaining components
of $\mathcal{Z}$ in turn: platform pressure ($\mathrm{Interv}_{u,t}$), then campaign
coordination ($\mathbf{c}_{C(u),t}$), then the account's own trajectory
($\mathbf{h}_{u,<t}$). Together they account for all of $\mathcal{Z}$. The state of an
account on a given tick is thus a function of the base parameters fixed at instantiation,
the platform's interventions, the directives imposed by the campaign, and the account's
history up to that point.

\subsubsection{The rollout}

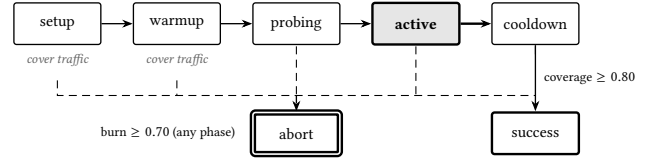
\begin{figure}[t]
\centering
\begin{tikzpicture}[x=0.93cm, y=1cm,
  ph/.style={box, minimum width=1.15cm, minimum height=0.55cm}]
  \node[ph] (setup)  at (0,0)    {setup};
  \node[ph] (warm)   at (1.7,0)  {warmup};
  \node[ph] (probe)  at (3.4,0)  {probing};
  \node[ph, fill=black!10, line width=0.9pt] (active) at (5.1,0) {\textbf{active}};
  \node[ph] (cool)   at (6.8,0)  {cooldown};
  \node[ph, line width=0.9pt] (succ) at (6.8,-1.45) {success};
  \node[ph, line width=0.9pt, double] (abort) at (3.4,-1.45) {abort};
  \node[note] at (0,-0.48) {cover traffic};
  \node[note] at (1.7,-0.48) {cover traffic};
  \draw[flow] (setup) -- (warm);
  \draw[flow] (warm) -- (probe);
  \draw[flow] (probe) -- (active);
  \draw[flow, line width=0.9pt] (active) -- (cool);
  \draw[flow] (cool) -- node[right, font=\tiny] {coverage $\ge 0.80$} (succ);
  \draw[densely dashed] (0,-0.75) -- (0,-0.95) -- (6.8,-0.95);
  \draw[densely dashed] (1.7,-0.75) -- (1.7,-0.95);
  \foreach \x in {3.4,5.1} \draw[densely dashed] (\x,-0.3) -- (\x,-0.95);
  \draw[densely dashed] (6.8,-0.3) -- (6.8,-0.45);
  \draw[flow, densely dashed] (3.4,-0.95) -- (abort);
  \node[font=\tiny, anchor=east, xshift=-3pt] at (abort.west) {burn $\ge 0.70$ (any phase)};
\end{tikzpicture}
\caption{Distillation campaign lifecycle. Setup and warmup emit benign cover traffic,
making the account indistinguishable from a casual user; the systematic extraction pattern
only emerges from probing onward. Abort can fire from any phase once $70\%$ of the fleet is
burned.}
\label{fig:phase}
\Description{Phase state machine: setup, warmup, probing, active, cooldown, then success when coverage reaches 0.80; abort from any phase when 70 percent of the fleet is burned.}
\end{figure}

With the campaign instantiated, the simulator rolls it out tick by tick, emitting
user-session records. The rollout is driven by the phase schedule
(Figure~\ref{fig:phase}).

% ----------------------------------------------------------- FIGURE 5 ----

Time advances in discrete daily ticks. On each tick, before any session is drawn, the
campaign performs five bookkeeping steps in a fixed order, each reading the state written
by the previous one.

\emph{(1) Decay.} Interventions imposed earlier (a rate limit, a CAPTCHA gate, a period of
enhanced monitoring) carry a countdown; each timer is decremented and expired
interventions are lifted, allowing a penalised account to recover full throughput
gradually.

\emph{(2) Fallout.} Every account suspended since the last tick is marked burned and
removed from the active roster, and the campaign checks its survival condition: if the
cumulative burn fraction reaches $0.70$, it aborts. Otherwise, sophisticated campaigns
($s > 0.5$) redistribute the load of burned accounts across the survivors per
Eq.~\eqref{eq:mult}, holding total throughput roughly constant as the fleet shrinks; the
multiplier is capped at $3\times$ so that survivors do not reveal themselves through
implausible activity spikes. Unsophisticated campaigns skip this step and simply lose
throughput.

\emph{(3) Phase transition.} The campaign consults its schedule and coverage progress to
decide whether to advance, e.g.\ from probing to active extraction.

\emph{(4) Shared-state update.} The campaign aggregates the domain coverage achieved by its
accounts into a single progress measure:
\begin{equation}
\text{cov} = \frac{\sum_{u \in \text{active distill.}} \text{covered\_slots}_u}{\text{total\_domain\_sectors}}.
\end{equation}
This quantity is read by the next tick's success check ($\geq 0.80$) and also signals
to each account how much of the domain remains uncovered, which shapes its activity.

\emph{(5) Activity allocation.} The campaign issues a daily directive to each surviving
account. For a distillation account in the active phase, the key field is the number of new
domain slots to target that day:
\begin{equation}
\text{slots\_today} = \min\big(\text{slots remaining in block},\ \operatorname{Poi}(2) + 1\big).
\end{equation}

Only after this bookkeeping do accounts interact with the model. For each active account,
the rollout assembles a sampling context, which summarises everything the generative model
needs about that account on that day, and then draws sessions one at a time. The context
combines three sources, each carrying a different aspect of the campaign mechanics
(Table~\ref{tab:context}).

\begin{table}[t]
\caption{The three influences stitched into each day's sampling context.}
\label{tab:context}
\footnotesize
\begin{tabularx}{\columnwidth}{@{}>{\raggedright\arraybackslash}p{0.19\columnwidth}
                                   >{\raggedright\arraybackslash}X
                                   >{\raggedright\arraybackslash}X@{}}
\toprule
\textbf{Influence} & \textbf{Captures} & \textbf{What it carries for distillation} \\
\midrule
Intervention state & what the platform is currently doing to the account
  & whether it is rate-limited; its sophistication (which governs how it evades) \\
Campaign context   & the cross-account coordination signals
  & today's slot allocation; remaining coverage, $1 - \text{cov}$ \\
Trajectory         & the account's own accumulated history
  & its personal saturation, $\text{covered}/\text{assigned} \in [0,1]$ \\
\bottomrule
\end{tabularx}
\end{table}

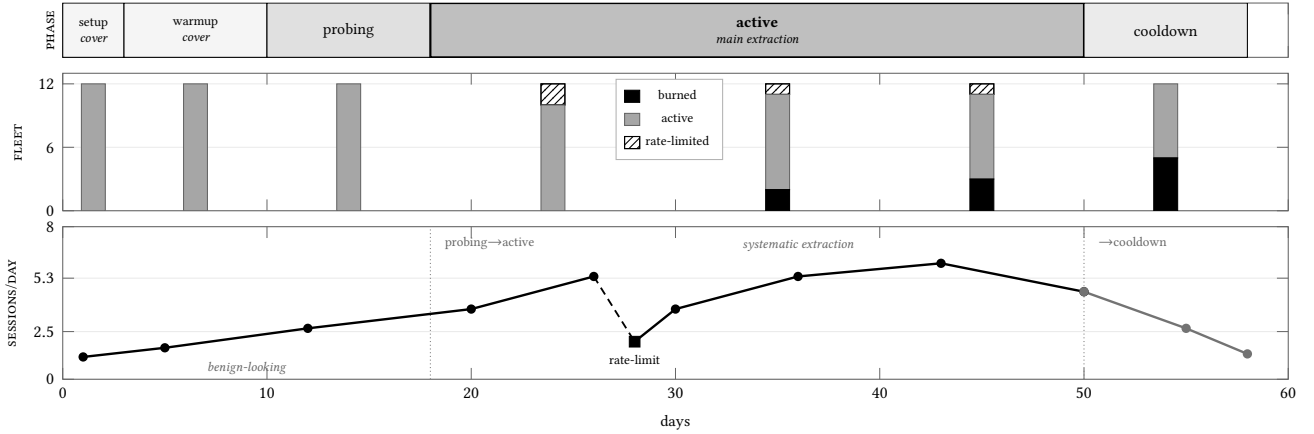
\begin{figure*}[t]
\centering
\begin{tikzpicture}
\begin{groupplot}[
  group style={group size=1 by 3, vertical sep=6pt, x descriptions at=edge bottom},
  width=\textwidth, xmin=0, xmax=60,
  xtick={0,10,...,60}, xlabel={days},
  tick label style={font=\scriptsize}, label style={font=\scriptsize},
  axis line style={black!60}, tick style={black!60},
  ymajorgrids, grid style={black!8},
  ylabel style={font=\scriptsize\scshape},
]
% --- phases
\nextgroupplot[height=2.3cm, ymin=0, ymax=1, ytick=\empty, ylabel={phase}, ymajorgrids=false]
  \draw[fill=black!4]  (axis cs:0,0)  rectangle (axis cs:3,1);
  \draw[fill=black!4]  (axis cs:3,0)  rectangle (axis cs:10,1);
  \draw[fill=black!12] (axis cs:10,0) rectangle (axis cs:18,1);
  \draw[fill=black!25, line width=0.8pt] (axis cs:18,0) rectangle (axis cs:50,1);
  \draw[fill=black!8]  (axis cs:50,0) rectangle (axis cs:58,1);
  \node[font=\tiny, align=center] at (axis cs:1.5,0.5) {setup\\\itshape cover};
  \node[font=\tiny, align=center] at (axis cs:6.5,0.5) {warmup\\\itshape cover};
  \node[font=\scriptsize] at (axis cs:14,0.5) {probing};
  \node[font=\scriptsize, align=center] at (axis cs:34,0.5) {\textbf{active}\\[-1pt]{\tiny\itshape main extraction}};
  \node[font=\scriptsize] at (axis cs:54,0.5) {cooldown};
% --- fleet
\nextgroupplot[height=3.4cm, ymin=0, ymax=13, ytick={0,6,12}, ylabel={fleet},
  ybar stacked, bar width=9pt,
  legend style={font=\tiny, at={(0.495,0.96)}, anchor=north, draw=black!30,
                legend columns=1},
  legend image code/.code={\draw[#1] (0cm,-0.08cm) rectangle (0.2cm,0.08cm);}]
  \addplot[fill=black, draw=black] coordinates {(1.5,0) (6.5,0) (14,0) (24,0) (35,2) (45,3) (54,5)};
  \addplot[fill=black!35, draw=black!60] coordinates {(1.5,12) (6.5,12) (14,12) (24,10) (35,9) (45,8) (54,7)};
  \addplot[fill=white, draw=black, postaction={pattern=north east lines}] coordinates {(1.5,0) (6.5,0) (14,0) (24,2) (35,1) (45,1) (54,0)};
  \legend{burned, active, rate-limited}
% --- sessions / day
\nextgroupplot[height=3.6cm, ymin=0, ymax=8, ytick={0,2.5,5.3,8}, ylabel={sessions/day}]
  \draw[densely dotted, black!50] (axis cs:18,0) -- (axis cs:18,8);
  \draw[densely dotted, black!50] (axis cs:50,0) -- (axis cs:50,8);
  \node[font=\tiny, anchor=north west, text=black!60] at (axis cs:18.3,7.9) {probing$\to$active};
  \node[font=\tiny, anchor=north west, text=black!60] at (axis cs:50.3,7.9) {$\to$cooldown};
  \addplot[black, line width=0.9pt, mark=*, mark size=1.3pt]
    coordinates {(1,1.17) (5,1.65) (12,2.67) (20,3.68) (26,5.39)};
  \addplot[black, densely dashed, line width=0.7pt] coordinates {(26,5.39) (28,1.97)};
  \addplot[black, line width=0.9pt, mark=*, mark size=1.3pt]
    coordinates {(28,1.97) (30,3.68) (36,5.39) (43,6.08) (50,4.59)};
  \addplot[black!55, line width=0.9pt, mark=*, mark size=1.3pt]
    coordinates {(50,4.59) (55,2.67) (58,1.33)};
  \addplot[only marks, mark=square*, mark size=2pt, black] coordinates {(28,1.97)};
  \node[font=\tiny, anchor=north] at (axis cs:28,1.7) {rate-limit};
  \node[font=\tiny\itshape, text=black!60] at (axis cs:9,0.6) {benign-looking};
  \node[font=\tiny\itshape, text=black!60] at (axis cs:36,7.0) {systematic extraction};
\end{groupplot}
\end{tikzpicture}
\caption{A 60-day distillation campaign rollout. Top: campaign phase bands. Middle: fleet
composition per tick---active, rate-limited, and burned accounts. Bottom: a single
account's session-density trace, with a rate-limit event mid-campaign and the
characteristic ramp from a benign flatline to active extraction.}
\label{fig:journey}
\Description{Three aligned panels over 60 days: phase bands, a stacked bar chart of active, rate-limited and burned accounts, and a line of one account's sessions per day rising through the active phase with a dip at a rate-limit event.}
\end{figure*}

Crucially, sessions within a day are not drawn independently: they are generated
sequentially, and the account's cumulative statistics are updated between draws. After
each session, the newly covered slots are recorded,
\begin{equation}
\text{covered\_slots} \mathrel{+}= \max\big(1,\ \lfloor \delta \cdot \text{assigned\_slots} \rfloor\big),
\end{equation}
so the next session on the same day already sees a slightly more saturated trajectory. This
ensures that an account does not cover the same territory twice and that its progress
tapers naturally as its block fills. Because these records are exactly what the campaign
aggregates at the start of the next tick, the loop is closed: each day's sessions determine
the next day's progress measure and directives, and the campaign advances until it either
covers $80\%$ of the domain or is aborted (Figure~\ref{fig:journey}).

% ----------------------------------------------------------- FIGURE 6 ----

\subsection{The Dataset}

The public dataset~\cite{daoistdurian2026dataset} is \fiveel's raw service-usage log for
its frontier models (model endpoints are named after the Claude family for simplicity). A
stratified sample of rows, one per attack type, is given in Appendix~\ref{app:sample}.
Campaign types are distinct from the attack type of each individual account within a
campaign. Several columns come in two variants (e.g.\ \cls{noisy\_avg\_prompt\_perplexity}
and \cls{avg\_prompt\_perplexity}): the former is the realistic noise-injected signal and
the latter its oracle value. The dataset has three notable simplifications:
\begin{enumerate}
  \item \textbf{Fixed roles.} An adversarial account is locked to a single campaign and
    attack type for its entire life; no \cls{harmful\_use} account later turns to
    \cls{distillation}, or vice versa. Real operators do pivot surfaces mid-campaign; we
    fix roles for tractability.
  \item \textbf{Cold start.} The fleet spawns directly into \cls{setup} with no burn-in,
    so the first days of any run are a transient rather than a steady state. A more
    faithful simulation would run a longer horizon and discard the initial period; we
    handle this at training time instead (Section~\ref{sec:features}).
  \item \textbf{Label delay is a prior, not a mechanism.} Each surface draws its arrival
    lag from its own log-normal delay distribution (\cls{dos} resolves in under a day,
    \cls{distillation} has a ${\sim}90$-day median), so at the type level delay tracks how
    hard a surface is to catch. The medians were set by hand to encode the intuition,
    familiar to T\&S teams, that conspicuous attacks resolve quickly while quiet, patient
    ones resolve slowly. There is, however, no account-level causality beneath this prior:
    two \cls{distillation} accounts draw from the same distribution regardless of their
    \cls{sophistication} or how convincingly either impersonated a power user.
\end{enumerate}
These simplifications make the synthetic task easier than its real counterpart, but the
dataset remains sufficient for the questions studied here.

% ============================================================= TRAINING ==
\section{Training the Detection Model}
\label{sec:training}

Recall from Section~\ref{sec:setup} that, given a session feature vector
$\mathbf{x} \in \mathbb{R}^d$, we want to predict two targets: the binary
$y_{\text{adv}} \in \{0,1\}$ (benign vs.\ adversarial) and the multi-class
$y_{\text{type}} \in \mathcal{C}$ (which kind of adversarial account). Here $\mathcal{C}$
corresponds to \cls{attack\_type} (not \cls{campaign\_type}); that is, we detect the type
of adversariality of a \emph{given account}, not of the campaign \emph{the account belongs
to}. The former is the simpler task and a natural starting point; the dataset
accordingly focuses on single-attack campaigns, with the exception
of \cls{multi\_surface} campaigns, which are an ensemble of attack types. Since
$y_{\text{type}}$ is strictly more descriptive than $y_{\text{adv}}$, we focus on the
multi-class case.

\subsection{Labels and Observability}
\label{sec:labels}

The raw data contains several label columns, such as the \cls{campaign\_type},
\cls{campaign\_label}, \cls{noisy\_campaign\_label} triplet. These labels vary along
\textbf{two independent axes}: their \textbf{purpose}, i.e.\ \emph{what} a label describes,
and their \textbf{observability}, i.e.\ how faithfully the channel that produced the label
reflects the ground truth.

\paragraph{Axis 1, purpose.} Three label families, nested from coarse to fine:
\begin{enumerate}
  \item \cls{is\_adversarial}, \emph{the binary verdict}: 1 for an adversarial account of
    any kind, 0 for an honest one. This is the target of the binary task, but it carries no
    information beyond the type label: it is a collapse of \cls{attack\_type}, with
    everything outside the \cls{benign.*} archetypes mapped to 1.
  \item \cls{attack\_type}, \emph{the per-account type}: the role of the account in a
    campaign: \cls{benign.\{casual\allowbreak|power\_user\allowbreak|developer\allowbreak|enterprise\allowbreak|researcher\}}, or one of
    the adversarial types, plus \cls{cover\_traffic}. Each adversarial account serves
    exactly one vector (the fixed-role assumption), and this is the target of the
    multi-class model. During \cls{warmup}/\cls{setup} an adversarial account behaves as a
    benign archetype, since it has not yet begun misbehaving, and within a
    \cls{multi\_surface} ring an account can play a \cls{cover\_traffic} role that
    differs from the ring's actual objective.
  \item \cls{campaign\_type}, \emph{the per-ring type}: the surface of the coordinated
    operation the account belongs to, absent for organic benign accounts. Because several
    accounts map to one campaign, this is a \emph{group} property; it is out of scope for
    the per-session predictor but retained so that a higher-rung model can aggregate across
    accounts and recover the ring.
\end{enumerate}

\paragraph{Axis 2, observability.} Here synthetic data offers a unique advantage. In
practice a label is never observed directly or perfectly, but through an evidentiary
channel that is delayed, lossy, or wrong. We model three levels, \M{3}, \M{2} and \M{1}, in
\emph{decreasing} order of fidelity and, correspondingly, \emph{increasing} order of
realism:
\begin{enumerate}
  \item \textbf{\M{3}, oracle.} The ground truth known to the simulator by construction:
    \cls{attack\_type}, \cls{campaign\_type}, \cls{is\_adversarial}, the exact
    \cls{attack\_phase}, and the latent \cls{sophistication}, available the instant a
    session is sampled, with no noise, delay, or ambiguity. \M{3} does not exist in a real
    T\&S setting and is used for evaluation only.
  \item \textbf{\M{2}, investigation-confirmed (``golden'').} The latent fields are removed
    and only the \emph{coarse} labels are kept, subject to a realistic
    \textbf{investigation delay} $\tau \sim \operatorname{LogNormal}(\mu_d, \sigma_d)$,
    typically 14--90 days between a session and an analyst's confirmed verdict. \M{2} is
    given by the \cls{label} / \cls{campaign\_label} columns. It corresponds to the golden
    dataset a mature T\&S team accumulates over years: clean and trustworthy, but late.
    It is also used for evaluation only.
  \item \textbf{\M{1}, operational.} The investigation channel is removed as well, leaving
    only the \emph{policy outputs of the system itself} (whether an account was suspended)
    plus a label-noise model: false positives from over-eager policies, false negatives
    from attackers who exfiltrate and leave before any rule fires, misattribution, and the
    same delay. \M{1} is given by \cls{noisy\_label} / \cls{noisy\_campaign\_label}. It is
    what a T\&S team holds most of the time, and the only channel that scales without
    consuming analyst time. \textbf{We train on \M{1}.}
\end{enumerate}
Table~\ref{tab:labelgrid} summarises the full labelling scheme.

\begin{table}[t]
\caption{The labelling scheme: purpose $\times$ observability.}
\label{tab:labelgrid}
\footnotesize
\setlength{\tabcolsep}{4pt}
\begin{tabular}{@{}llll@{}}
\toprule
\textbf{Channel} & \textbf{Binary} & \textbf{Attack type} & \textbf{Campaign type} \\
                 &                 & \emph{(per account)} & \emph{(per ring)} \\
\midrule
\M{3} oracle      & \cls{is\_adversarial} & \cls{attack\_type} & \cls{campaign\_type} \\
\M{2} golden      & \emph{(derived)} & \cls{label}        & \cls{campaign\_label} \\
\M{1} operational & \emph{(derived)} & \cls{noisy\_label} & \cls{noisy\_campaign\_label} \\
\bottomrule
\multicolumn{4}{@{}l}{\tiny \M{3}: no delay, no noise; \M{2}: + investigation delay; \M{1}: + delay + noise.}
\end{tabular}
\end{table}

\paragraph{The information-loss hierarchy.} Each row is a \emph{stochastic function of the
row above it}: \M{2} coarsens and delays \M{3}; \M{1} corrupts and delays \M{2}. Having all
three lets us evaluate the model in different settings. Evaluation against \M{3} gives the
\textbf{ceiling}, the irreducible Bayes error once the world is fully observed. Evaluation
against \M{2} indicates what a \textbf{well-resourced} T\&S team could plausibly achieve.
Evaluation against \M{1} indicates what can be \textbf{deployed today} with the labels
actually available, and, by subtraction, how much performance is lost by not investing in
a labelling pipeline. We report results against all three.

\paragraph{How the noise and delay are applied.} Generating \M{1} by flipping labels
i.i.d.\ per session would be unrealistic. Once a single session of an account is
annotated, that verdict \emph{recontextualises the entire account}: an analyst does not
conclude that one session was a jailbreak and the next was benign, but that \emph{the
account} is malicious, and that judgement applies to its entire history. Per-session
corruption would destroy exactly the cross-session consistency that a real labelling
process imposes. We therefore corrupt labels with an \textbf{account-level confusion
draw}: \emph{one} draw per account, not per session, from a confusion
distribution (hand-tuned per surface, in the same spirit as the log-normal delay), deciding
whether and how that account's labels are corrupted: benign flagged as adversarial (FP),
adversarial missed entirely (FN), or one attack type misread as another
(misattribution). The outcome of the draw is applied \emph{consistently} across every
session of the account.

The counterpart of consistent noise is \textbf{consistent backfilling}. When an account is
confirmed malicious, a T\&S team does not only tag the session that triggered the alert;
it reviews and relabels the account's history. \emph{How far} back it relabels is not
uniform, and handling this correctly keeps the dataset both realistic and free of temporal
leakage. The deciding factor is whether an attack type is \emph{always-adversarial},
malicious from its first session (e.g.\ \cls{dos} and \cls{distillation}), or can
plausibly appear innocent early on (e.g.\ a \cls{jailbreak} prober who initially resembles
a curious researcher). The generator follows the rules in Table~\ref{tab:backfill}.

\begin{table}[t]
\caption{Backfilling rules by account fate and attack type.}
\label{tab:backfill}
\footnotesize
\begin{tabularx}{\columnwidth}{@{}>{\raggedright\arraybackslash}p{0.2\columnwidth}
                                   >{\raggedright\arraybackslash}p{0.2\columnwidth}
                                   >{\raggedright\arraybackslash}X@{}}
\toprule
\textbf{Account fate} & \textbf{Type} & \textbf{What gets labelled} \\
\midrule
Suspended / revoked      & always-adv. & \textbf{All} sessions backfilled---intentional from day one \\
Suspended / revoked      & not always-adv. & \textbf{Only} the session that triggered suspension---earlier ones were genuinely ambiguous \\
Active, label lands late & always-adv. & \textbf{All} sessions backfilled once the investigation retroactively confirms them \\
Active, label lands late & not always-adv. & Only sessions from \cls{label\_arrival\_day} onward---preserving temporal honesty \\
Never confirmed          & any & \textbf{Unlabelled}---the adversaries that simply slip through \\
\bottomrule
\end{tabularx}
\end{table}

The purpose of this split is \textbf{temporal integrity}. An always-adversarial account
was malicious \emph{by construction} from its first day, so backfilling its entire history
leaks nothing. A not-always-adversarial account, however, genuinely \emph{was} ambiguous
early on, so it is labelled only from the confirmation day forward; labelling it from day
one would leak information from the future and inflate every offline metric.

\subsection{Feature Engineering}
\label{sec:features}

The raw data is a session-level log with dozens of columns, oracle and noisy signals side
by side, and benign and adversarial traffic interleaved. We transform it into a clean
numeric matrix $\mathbf{X}$ suitable for gradient-boosted trees in four steps
(Table~\ref{tab:pipeline}).

\paragraph{Step 1: clip the horizon at both ends.} Every account spawns directly into
\cls{setup} with no burn-in, so the opening ticks are a cold-start transient rather than a
steady state. The tail has the opposite defect: for a session near the final day, the
14-to-90-day investigation lag has not yet elapsed, so its \M{1}/\M{2} labels read
\cls{benign} only because the verdict has not arrived, a right-censoring artefact of the
delay. We therefore discard a burn-in head and a burn-out tail and keep the
steady-state band $n_{\text{burn\_in}} \le \text{day} \le n_{\text{burn\_out}}$, with
$n_{\text{burn\_in}} = 40$ and $n_{\text{burn\_out}} = 320$. The head clip removes the
warm-up transient; the tail clip acts as a \emph{label-delay backstop}, preventing training
on sessions whose \cls{benign} label reflects only data immaturity.

\paragraph{Step 2: materialise the label grid.} The \M{3}$\to$\M{2}$\to$\M{1} hierarchy
becomes nine concrete columns, one triplet (\cls{is\_adversarial}, \cls{attack\_type},
\cls{campaign\_type}) per channel. The construction rule is uniform: a null in the source
label means ``no confirmed adversariality on this channel'', so \cls{is\_adversarial} is an
\texttt{is\_not\_null} cast to an integer, while the type strings get their nulls filled
with the literal \cls{benign}.

\paragraph{Step 3: compute per-account rolling history features.} A single session carries
little information in isolation; the discriminating signal lies in the
\emph{trajectory}. A bot farm's refusal count rises day over day, a distillation ring's
domain coverage grows steadily, and a jailbreak prober's safety scores drift as it warms
up. To expose this temporal structure to the trees without a recurrent model, we summarise
each account's recent history over a trailing seven-day window,
\begin{equation}
W^{(7)}_{u,t} = \big\{\, s : u_s = u,\ \ t - 7 < d_s \le t \,\big\},
\end{equation}
the set of sessions of account $u$ in the seven days ending on (and inclusive of) day $t$.
Each rolling feature is a simple aggregation over that window,
\begin{equation}
\operatorname{mean}_{u,t}(x) = \frac{1}{|W^{(7)}_{u,t}|}\sum_{s \in W^{(7)}_{u,t}} x_s,
\qquad
\operatorname{sum}_{u,t}(x) = \sum_{s \in W^{(7)}_{u,t}} x_s,
\end{equation}
and likewise for running maxima and counts, applied to a handful of base signals: session
volume, average request and token counts, the noisy safety / timing / coverage
measurements, accumulated refusals, and intervention histories. The key constraint is
$d_s \le t$: the window is \emph{strictly trailing} and never includes a session from the
future.

\paragraph{Step 4: select the allowed feature set.} The exclusions here matter as much as
the inclusions. We group the observable columns into tiers according to how much
processing separates the raw event from the value:
\begin{enumerate}
  \item \textbf{Tier 1, fully observable.} Counts and facts the platform logs
    directly: \cls{n\_requests}, \cls{session\_duration\_min}, \cls{total\_tokens},
    \cls{output\_input\_ratio}, inter-request timing mean/std, the \cls{is\_api} /
    \cls{is\_datacenter\_ip} / \cls{is\_vpn} flags, and \cls{n\_refusals}.
  \item \textbf{Tier 2, model-derived.} Signals that are themselves outputs of an upstream
    model: prompt perplexity, safety scores, query-structure entropy, tool-use rate, and
    domain-coverage delta. We use the \textbf{noisy} variants, because in production the
    defender only sees the noised measurement.
  \item \textbf{Tier 3, derived / aggregated.} Cross-account graph signals such as
    \cls{noisy\_shared\_ip\_count}, \cls{noisy\_shared\_device\_count},
    \cls{registration\_velocity}, and \cls{concurrent\_sessions}, again in their noisy
    variants.
\end{enumerate}
The family of surface flags (\cls{template\_query\_pattern},
\cls{divergence\_probe\_pattern}, \cls{token\_flood\_pattern}, and others) is excluded
entirely: in our generator these are close to oracle signals and are unlikely to be
available in practice without substantial preprocessing, although they remain useful for
diagnostics. A real T\&S team holds the \emph{raw} perplexity and entropy signals and must
\emph{learn} the boundary that a surface flag hard-codes; we hold our model to the same
standard.

\begin{table*}[t]
\caption{The feature-engineering pipeline at a glance.}
\label{tab:pipeline}
\small
\begin{tabularx}{\textwidth}{@{}c l >{\raggedright\arraybackslash}X >{\raggedright\arraybackslash}X@{}}
\toprule
\textbf{Step} & \textbf{Transformation} & \textbf{What it does} & \textbf{Why} \\
\midrule
1 & Clip horizon & Keep the burn-in/burn-out band ($n_{\text{burn\_in}}{=}40$ to $n_{\text{burn\_out}}{=}320$)
  & Drop the cold-start transient (head) and the censored tail whose labels have not arrived \\
2 & Materialise label grid & Fold the \cls{noisy\_*} / \cls{label} / \cls{attack\_type} channels into \M{1}/\M{2}/\M{3} $\times$ (\cls{is\_adversarial}, \cls{attack\_type}, \cls{campaign\_type})
  & Turn the \M{3}$\to$\M{2}$\to$\M{1} hierarchy into concrete train/eval columns \\
3 & Rolling features & Trailing 7-day per-account aggregates ($W^{(7)}_{u,t}$, strictly $d_s \le t$)
  & Expose behavioural \emph{trajectory} to the model without leaking the future \\
4 & Select allowed features & Tiers 1--3 (noisy) + geo + cumulative, plus the rolling block
  & Restrict to production-observable signals; \textbf{exclude oracle-like surface flags} \\
\bottomrule
\end{tabularx}
\end{table*}

\subsection{Training Setup}

\begin{table}[t]
\caption{Adversarial prevalence per split and label channel.}
\label{tab:prevalence}
\small
\begin{tabular}{@{}lccc@{}}
\toprule
\textbf{Split} & \textbf{\M{1} (operational)} & \textbf{\M{2} (golden)} & \textbf{\M{3} (oracle)} \\
\midrule
Train & 1.53\% & 0.74\% & 2.77\% \\
Val   & 1.35\% & 0.57\% & 2.61\% \\
Test  & 1.47\% & 0.70\% & 3.22\% \\
\bottomrule
\end{tabular}
\end{table}

After cleaning, the adversarial prevalence across the three splits is as given in
Table~\ref{tab:prevalence}.\footnote{\textbf{Split protocol.} Rows are not split at random,
which would let the same account (and, worse, the same \emph{campaign}) span train,
validation and test, leaking the answer. Instead we use a \textbf{time-population split}:
we partition by account so that no user appears in two splits, \emph{and} cut along the
time axis so that training data lies strictly in the past relative to validation and test.
Disjoint users imply disjoint campaigns.} The oracle channel \M{3} records the most
adversarial activity and the operational \M{1} somewhat less, since much of it is missed
or backfilled late; \M{2} is lowest of all, as its investigation channel confirms only a
fraction of accounts, and late. Whichever channel is used, however, fewer than one session
in sixty is adversarial: the dataset is highly imbalanced.

This degree of imbalance is familiar from fraud, payments, and intrusion detection. In the
\emph{binary} setting the standard remedies are well established: prevalence-sensitive
metrics such as AUPRC rather than accuracy~\cite{mlm2020auprc}, class reweighting via
XGBoost's \texttt{scale\_pos\_weight}~\cite{xgboostparams}, or losses that down-weight
easy majority examples, such as focal loss~\cite{arora2020focal}. All share the aim of
preventing the abundant negative class from overwhelming the rare positive one.

Our task, however, is \textbf{multi-class}, so this aim must be generalised from two
classes to nine. We fit an XGBoost~\cite{chen2016xgboost} model with the
\texttt{multi:softprob} objective, which outputs a full probability simplex
$\hat{\mathbf{y}} \in \Delta^{8}$ over the nine classes, and use \texttt{mlogloss}
(multi-class cross-entropy) for early stopping~\cite{xgboostparams}. Since the scalar
\texttt{scale\_pos\_weight} does not extend beyond two classes, we instead supply a
\textbf{per-sample weight vector} that reshapes the loss so the model does not
under-value the rare surfaces.

\paragraph{Weights are applied at training time only.} Sample weights change the loss, and
therefore the gradient, so that during fitting the optimiser cannot minimise its objective
by ignoring the rare classes. At \emph{evaluation} time the opposite is required: a
calibrated measure of performance on data as it arrives in production, where the rare
classes really are rare. The \textbf{training} stream therefore passes through a
\emph{weighted} iterator, while the \textbf{validation} and \textbf{test} streams pass
through an \emph{unweighted} iterator, so that \texttt{mlogloss} is estimated, and early
stopping triggered, on the true distribution. We derive the weights in three steps.

\emph{Step 1: inverse-frequency weighting.} Each class $k$ receives a weight inversely
proportional to its frequency, so that a class seen a thousand times less often is
weighted a thousand times more heavily per example. With a smoothing exponent $\gamma$,
\begin{equation}
w_k = \left(\frac{N}{K \cdot n_k}\right)^{\gamma}, \qquad N = \textstyle\sum_k n_k,\quad K = 9,
\end{equation}
where $n_k$ is the count of class $k$. The weight is attached to each row by indexing this
length-$K$ vector with the row's label, $w = \texttt{class\_weights}[y]$, a single gather
that is cheap enough to perform on the fly inside the streaming iterator.

\emph{Step 2: square-root smoothing, $\gamma = \tfrac{1}{2}$.} Setting $\gamma = 1$
recovers \emph{raw} inverse frequency, which is unstable under our skew: a surface
$10{,}000\times$ rarer than benign would receive a $10{,}000\times$ weight, allowing a
handful of examples from a single rare class to dominate every gradient update and causing
the loss to oscillate. The square root compresses this dynamic range, so the same
$10{,}000\times$ rarity yields a weight of roughly $100\times$. This trades a small amount
of minority emphasis for substantially more stable training.

\emph{Step 3: the benign clamp.} Plain inverse frequency has an undesirable side effect:
because benign is the \emph{majority} class, it receives the \emph{smallest} weight, and
once every rare surface has been up-weighted, the benign class can be effectively removed
from the gradient. A model trained this way no longer learns where the benign boundary
lies and produces false positives on any unusual behaviour, which is costly when benign
traffic exceeds $98\%$ and every false positive adds friction for a paying customer. We
therefore floor the benign weight at that of the heaviest non-benign class,
\begin{equation}
w_{\cls{benign}} \leftarrow \max\Big(w_{\cls{benign}},\ \max_{k \neq \cls{benign}} w_k\Big),
\end{equation}
ensuring that the correction toward rare types never goes so far that the model loses
track of \emph{normal} behaviour.

\begin{table}[t]
\caption{Per-class weights under the three weighting stages, for an illustrative
class distribution consistent with the ${\sim}1.5\%$ adversarial train prevalence.
Spread = max/min weight.}
\label{tab:weights}
\footnotesize
\setlength{\tabcolsep}{4pt}
\begin{tabular}{@{}lrrrr@{}}
\toprule
\textbf{Class} & $n_k$ & \textbf{Inv.\ freq.} & \textbf{+ clamp} & \textbf{+ clamp + $\sqrt{\cdot}$} \\
\midrule
\cls{benign}            & 985{,}000 & 0.11  & 158.7 & 12.6 \\
\cls{bot\_farm}         & 4{,}100   & 27.1  & 27.1  & 5.2 \\
\cls{distillation}      & 3{,}200   & 34.7  & 34.7  & 5.9 \\
\cls{jailbreak}         & 2{,}600   & 42.7  & 42.7  & 6.5 \\
\cls{harmful\_use}      & 1{,}400   & 79.4  & 79.4  & 8.9 \\
\cls{dos}               & 1{,}100   & 101.0 & 101.0 & 10.1 \\
\cls{agentic\_misuse}   & 1{,}000   & 111.1 & 111.1 & 10.5 \\
\cls{credential\_abuse} & 900       & 123.5 & 123.5 & 11.1 \\
\cls{data\_extraction}  & 700       & 158.7 & 158.7 & 12.6 \\
\midrule
\multicolumn{2}{@{}l}{Spread}      & $1407\times$ & $5.9\times$ & $2.4\times$ \\
\bottomrule
\end{tabular}
\end{table}

Why not set $\gamma = 1$ and balance the classes \emph{perfectly}? Full inverse frequency
would weight each session of the rarest surface as heavily as over a thousand benign ones,
and the training labels are the noisy, operational \M{1} channel, containing false
positives, missed attacks, and misattributions by design. Fitting so aggressively on so
few rows leads the model to \emph{memorise} rather than \emph{learn} the rare surface,
including its sampling artefacts and mislabels, amplifying label noise. The $\sqrt{\cdot}$
smoothing and the clamp therefore serve a single purpose: not to \emph{equalise} the
classes, but to make the rare ones \emph{learnable} without allowing a handful of noisy
examples, or one over-weighted class, to dominate the fit.

Table~\ref{tab:weights} compares the three stages on one fixed, imbalanced class
distribution. \textbf{Plain inverse frequency} starves benign: its weight is near zero
while the rarest surfaces receive very large weights, a spread of more than a
thousandfold. \textbf{Adding the clamp} restores benign, raising its weight to match the
heaviest surface. This is a heuristic design choice; it is intuitively motivated and
performs well empirically, and we leave its theoretical grounding to future work.
Finally, \textbf{adding $\sqrt{\cdot}$ smoothing} flattens the distribution further, so
that no single surface dominates the gradient and training remains stable.

% ============================================================== RESULTS ==
\section{Results and Evaluation}
\label{sec:results}

We now evaluate whether the trained model is accurate and whether it is useful to the T\&S
team of \fiveel.

\subsection{Binary Detection}

\begin{table}[t]
\caption{Binary \cls{is\_malicious} evaluation on the test split, by label channel.}
\label{tab:binary}
\footnotesize
\setlength{\tabcolsep}{3.5pt}
\begin{tabular}{@{}lccccc@{}}
\toprule
\textbf{Channel} & \textbf{Prev.} & \textbf{AUROC} & \textbf{AUPRC} & \textbf{P@1\%FPR} & \textbf{R@1\%FPR} \\
\midrule
\M{1} operational & 1.47\% & 0.842 & 0.313 & 0.360 & 0.379 \\
\M{2} golden      & 0.70\% & 0.994 & 0.609 & 0.360 & 0.801 \\
\M{3} oracle      & 3.22\% & 0.997 & 0.993 & 0.767 & 0.993 \\
\bottomrule
\end{tabular}
\end{table}

Table~\ref{tab:binary} reports binary detection performance on the test split for each
level of label observability. Against the oracle \M{3}, the task is very nearly solved: an
AUPRC of $0.99$ and a recall of $0.99$ at a $1\%$ false-positive rate, meaning the model
detects essentially every adversary while rarely flagging a legitimate
user.\footnote{The synthetic setting is cleaner than a real one, so the \M{3} results
should be read as an optimistic ceiling.} Against \M{1}, the label type a real T\&S team
most commonly holds, the \emph{same model} appears mediocre, with an AUPRC of $0.31$.
However, \M{1} is \M{3} observed through delayed, noisy, and partially missing verdicts, so
evaluation against \M{1} measures the quality of the \emph{labels} as much as that of the
model. The model is, in effect, more accurate than the operational labels used to grade
it, and practitioners evaluating against such labels risk substantially underestimating
their detectors.

The gap between the \M{1}, \M{2} and \M{3} results also quantifies the value of a golden
dataset, costly as it is to build, and confirms that gradient-boosted trees remain a
strong baseline for tabular detection tasks of this kind.

\subsection{Attack-Type Attribution with a Naive Policy}

Beyond detection, identifying the \emph{attack type} is key to applying the correct
intervention and routing each case to the appropriate team (Table~\ref{tab:classes}).
Table~\ref{tab:naive} gives multi-class results when the argmax of the class-probability
vector is used as a naive decision policy.

\begin{table}[t]
\caption{Multi-class attack-type evaluation under the naive argmax policy.}
\label{tab:naive}
\footnotesize
\setlength{\tabcolsep}{4pt}
\begin{tabular}{@{}lccccc@{}}
\toprule
\textbf{Channel} & \textbf{Top-1} & \textbf{Top-3} & \textbf{Macro-acc} & \textbf{Log-loss} & \textbf{Routing} \\
\midrule
\M{1} operational & 0.988 & 0.997 & 0.194 & 0.067 & 0.989 \\
\M{2} golden      & 0.997 & 1.000 & 0.244 & 0.021 & 0.997 \\
\M{3} oracle      & 0.983 & 0.998 & 0.247 & 0.058 & 0.983 \\
\bottomrule
\end{tabular}
\end{table}

The high Top-$N$ metrics contrast sharply with the low macro-accuracy. Top-1 and routing
accuracy of $0.98$ appear strong, but the argmax, while free of tunable thresholds, is
\emph{prior-sensitive}: when benign accounts for $98\%$ of traffic, the majority class wins
every close call, and a model can achieve high top-1 accuracy by predicting \cls{benign}
almost uniformly. The prevalence-agnostic macro-accuracy is therefore the more meaningful
measure, and it reveals weak performance, although the \M{1}--\M{2}--\M{3} observability
gap remains visible.

To understand this, we break the aggregate down by class (Table~\ref{tab:perclass}),
comparing each class's \emph{ranking} with its \emph{decision}. \textbf{PR-AUC} is computed
one-vs-rest and measures whether the model ranks a class's true members above all others;
\textbf{F1} is measured on the class predicted by the naive argmax policy. The classes fall
into three groups.

\emph{Good ranking, compatible with argmax.} For \cls{benign} (F1 $0.99$), \cls{dos}
($0.62$), \cls{bot\_farm} ($0.48$) and \cls{credential\_abuse} ($0.45$), ranking and
decision agree: the model separates these surfaces well \emph{and} the argmax selects them
reasonably often. These classes have conspicuous, distinctive signatures and could be acted
on immediately.

\emph{Good ranking, incompatible with argmax.} For \cls{distillation} (PR-AUC $0.79$,
argmax-F1 $0.00$) and \cls{jailbreak} (PR-AUC $0.88$, F1 $0.01$), the model ranks positives
well, yet the argmax decision almost never selects them. The model separates these classes,
but their probabilities cannot overcome the large benign prior to win the argmax.

\begin{figure*}[t]
\centering
\newcommand{\prpanel}[1]{\includegraphics[width=0.3\textwidth]{figures/gbdt_attack_type_pr_curve_#1.png}}
\prpanel{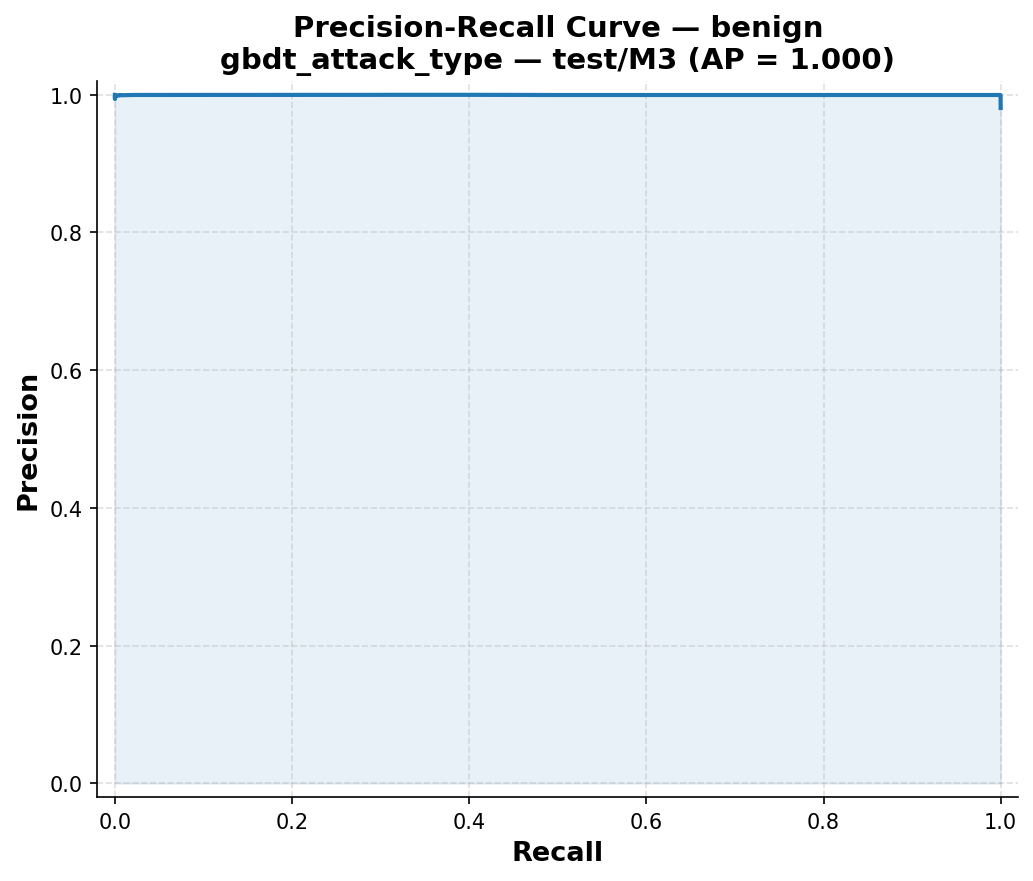}\hfill\prpanel{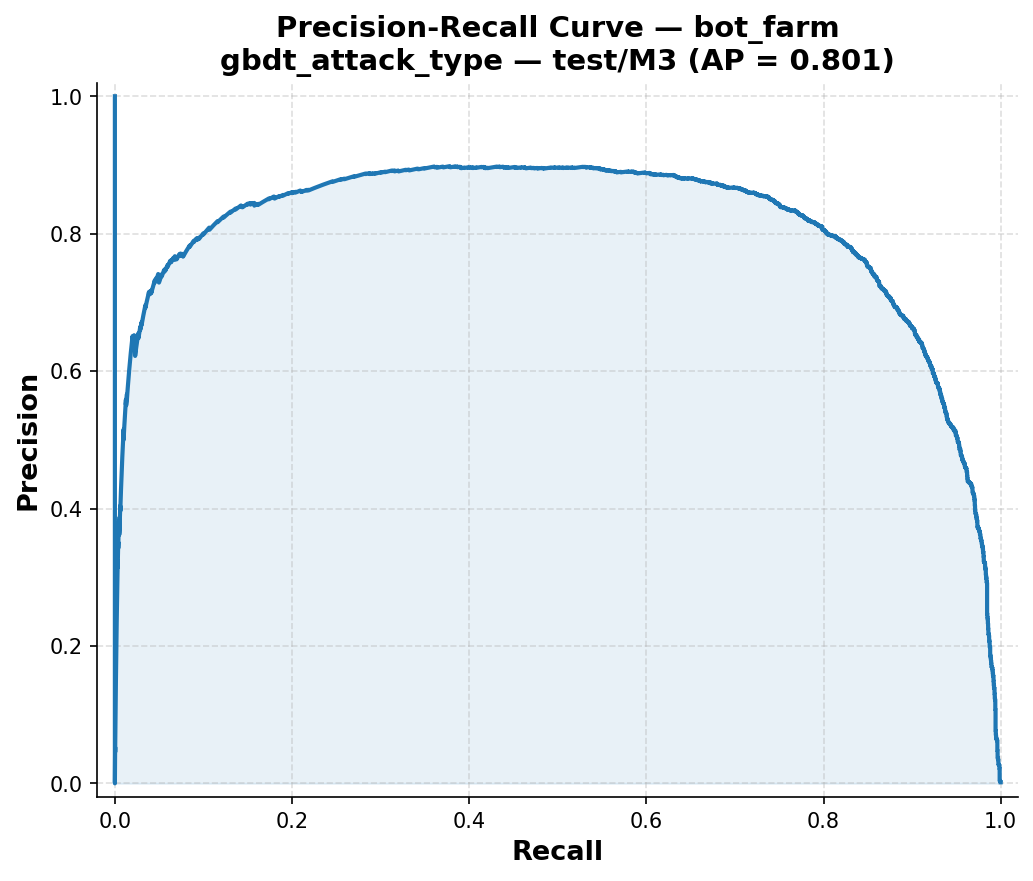}\hfill\prpanel{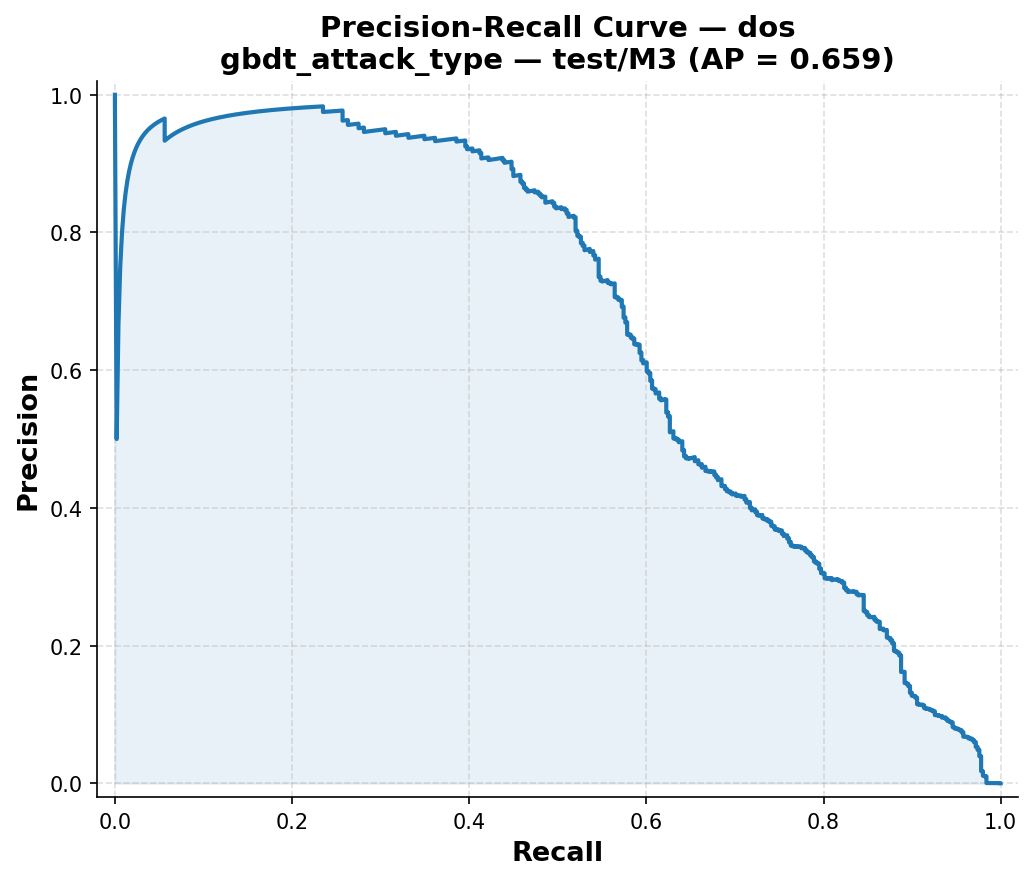}\\[4pt]
\prpanel{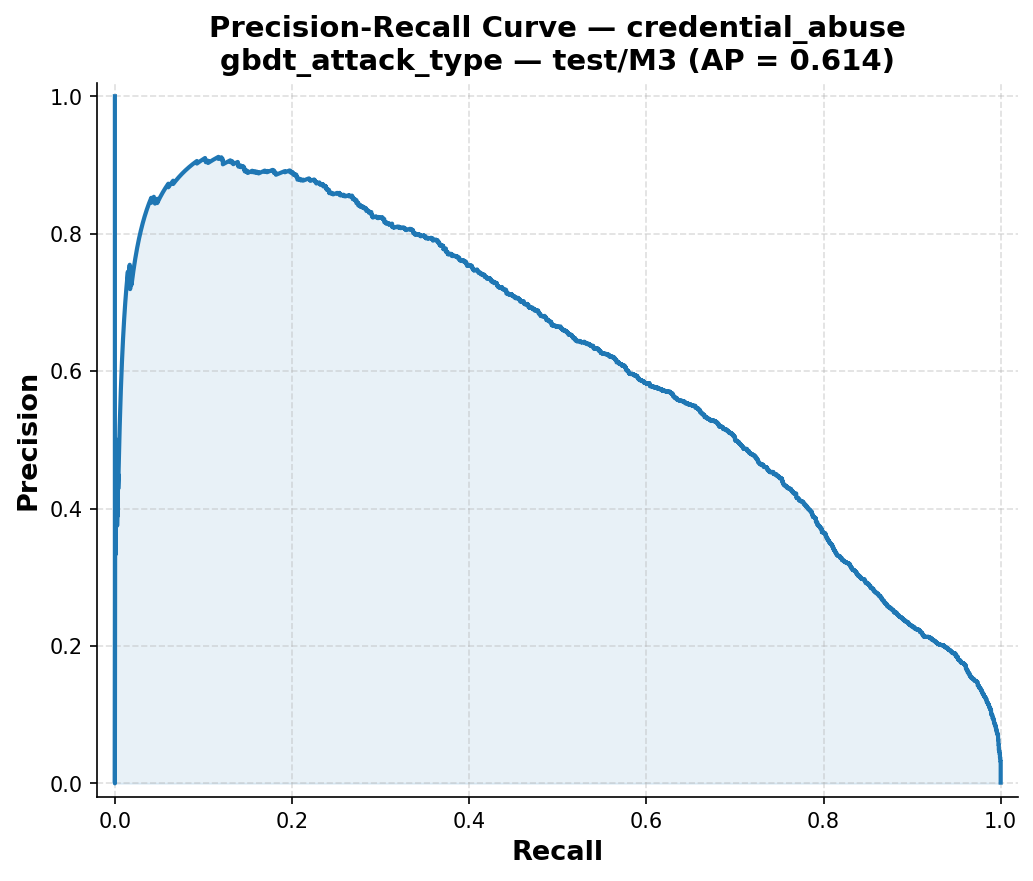}\hfill\prpanel{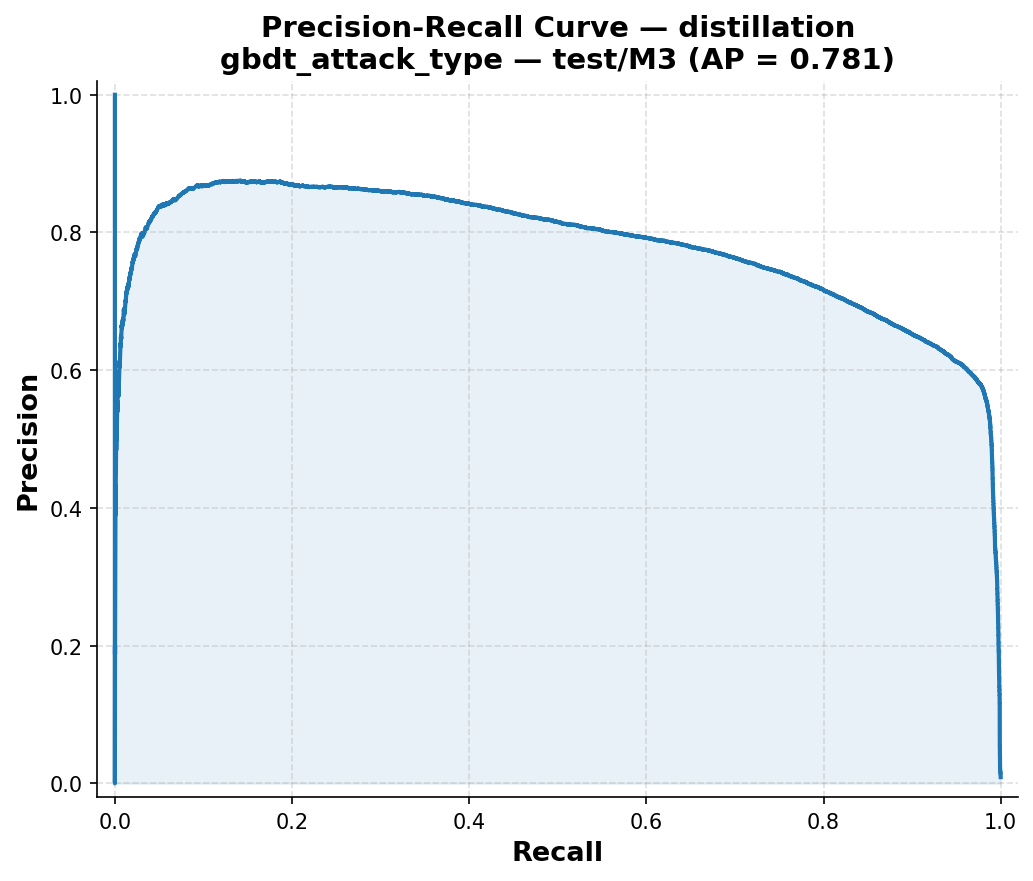}\hfill\prpanel{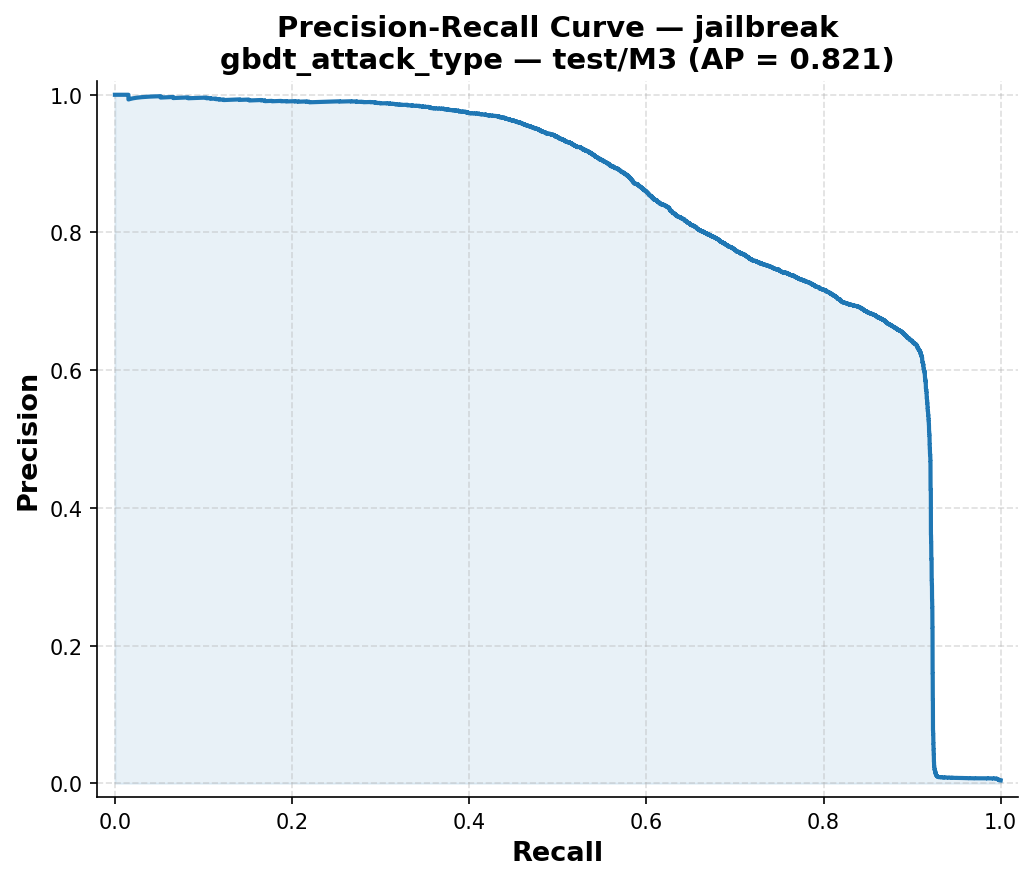}\\[4pt]
\prpanel{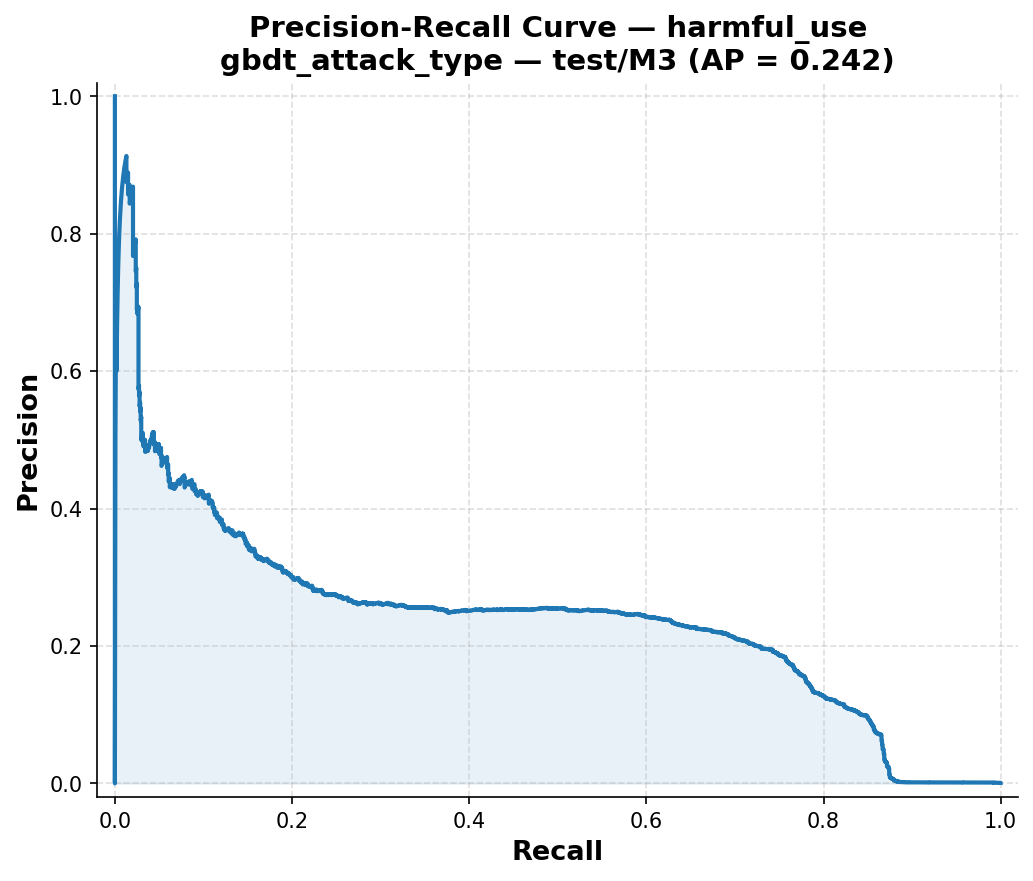}\hfill\prpanel{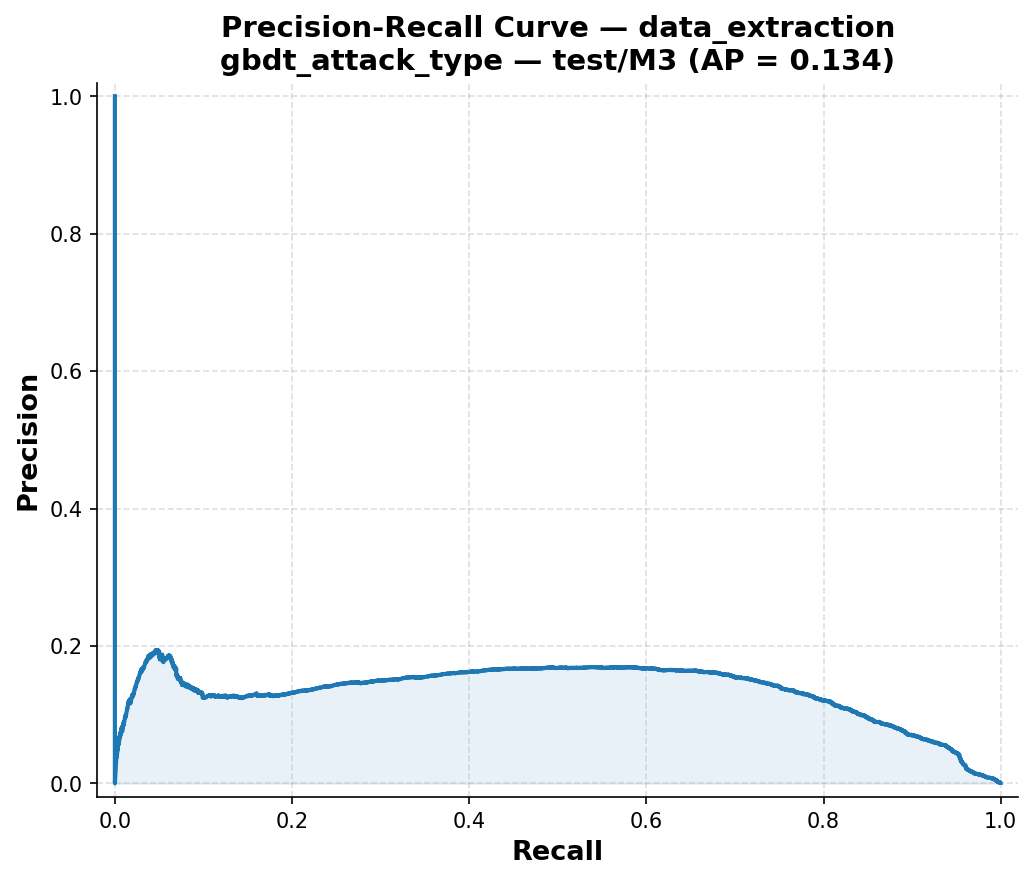}\hfill\prpanel{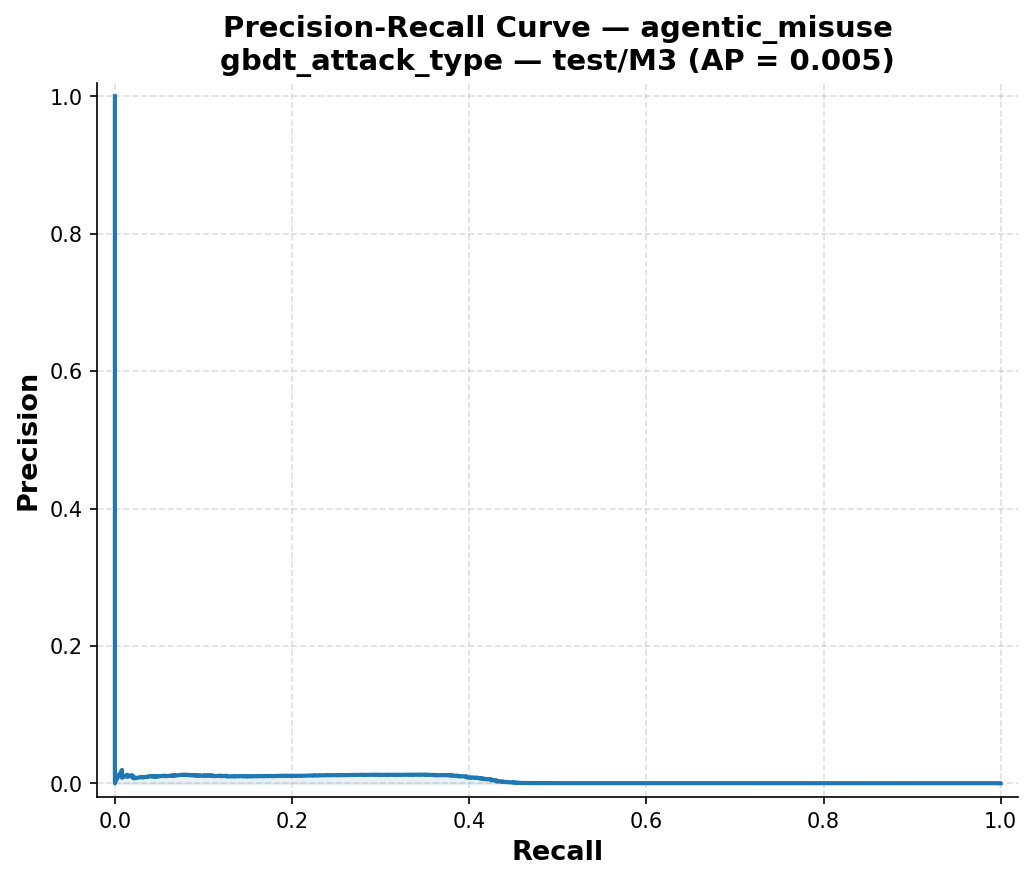}
\caption{Per-class precision--recall curves for the attack-type GBDT (test split,
\M{3}), ordered roughly best to worst. A curve that bows toward the top-right corner is
separable; note how \cls{distillation} and \cls{jailbreak} bow high even though their
argmax F1 sits at zero.}
\label{fig:prgrid}
\Description{Nine precision-recall curves, one per attack-type class.}
\end{figure*}

\emph{Weak ranking.} \cls{harmful\_use} (PR-AUC $0.25$), \cls{data\_extraction} ($0.14$)
and \cls{agentic\_misuse} ($0.01$) are poorly separated. Either the dataset contains too few
samples of these classes, or the engineered features fail to capture their signal. These
classes require more data or methods from higher rungs of the ladder, not a different
decision rule. Per-class precision--recall curves are given in
Figure~\ref{fig:prgrid}.

% ----------------------------------------------------------- FIGURE 7 ----

\subsection{Thresholded Decision Policies}

We replace the argmax with per-class decision thresholds, since each attack type has a
different prevalence and hence a different \emph{sensitivity}. This raises a new question:
when several classes clear their thresholds, which should be chosen? We address this with
a simple decision engine with two levers, which together generate a family of policies.

The first lever is the selection rule among contenders: the largest probability (an argmax
over a shortlist), or, among the classes that cleared their own threshold, the one with the
widest margin, measured either in absolute terms (the \emph{absolute lift}, $p_k - t_k$) or
relative to the threshold (the \emph{relative lift}, $p_k / t_k$, which is fairer when
thresholds differ greatly in magnitude). The second lever is the treatment of
\cls{benign}: it either competes like any other class or, as the overwhelming default, is
excluded unless it clears a markedly higher \textbf{floor}, so that suspect sessions receive
closer scrutiny.

As the best configuration is not known a priori, we evaluate six policies in addition to
the naive one. All thresholds are tuned on the validation set using the noisy \M{1}
labels. Let $p = (p_0, \dots, p_8)$ be the softmax vector over $\mathcal{C}$, $t_k$ the
tuned threshold for class $k$, and $b = 0$ the index of \cls{benign}. All policies depend on
the \textbf{contender set}, the classes that clear their own threshold:
\begin{equation}
\mathcal{K}(p) = \{\, k \in \mathcal{C} : p_k \ge t_k \,\}.
\end{equation}
Whenever a policy is left with an empty contender set, it abstains to \cls{benign}.

\paragraph{1. Absolute lift.} Every class competes, and the winner is the class that
exceeds its own threshold by the widest absolute margin:
$\hat{y} = \arg\max_{k \in \mathcal{K}(p)} (p_k - t_k)$. Because the margin is measured in
probability mass, a class with a high threshold needs a correspondingly high probability to
win, which implicitly favours classes the model is already confident about.

\paragraph{2. Relative lift.} As above, but the margin is scaled by the threshold, so
clearing a low threshold twice over beats clearing a high one narrowly:
$\hat{y} = \arg\max_{k \in \mathcal{K}(p)} p_k / t_k$. This comparison is fairer when
thresholds differ widely, and gives the rare classes, whose thresholds are necessarily low,
a realistic chance of selection.

\paragraph{3. Benign floor $\to$ argmax.} This policy uses the second lever: \cls{benign} is
no longer an ordinary contender and must clear a high floor $\tau_b = 0.65$. Otherwise it is
excluded, and the argmax is taken over the remaining classes:
\begin{equation}
\hat{y} = \begin{cases}
b, & p_b \ge \tau_b, \\
\arg\max_{k \in \mathcal{C} \setminus \{b\}} p_k, & \text{otherwise.}
\end{cases}
\end{equation}
This encodes an explicit prior in the form of a \emph{hard decision boundary} that shifts
decisions away from the majority class.

\paragraph{4. Benign floor $\to$ absolute lift.} As policy 3, except that once
\cls{benign} is excluded the remaining classes are ranked by absolute margin,
$\arg\max_{k \in \mathcal{K}(p) \setminus \{b\}} (p_k - t_k)$, with the floor tuned to
$\tau_b = 0.60$.

\paragraph{5. Benign floor $\to$ relative lift.} The most aggressive configuration:
\cls{benign} must be near-certain ($\tau_b = 0.95$) to be selected, and the remaining
contenders are ranked by relative margin,
$\arg\max_{k \in \mathcal{K}(p) \setminus \{b\}} p_k / t_k$. Both levers are at their
extremes: the majority class is held to a strict standard, while the minority classes are
compared on terms that do not penalise their rarity.

\paragraph{6. Meta-stacker.} The only learned policy: a smaller meta-model, here a simpler
XGBoost instance, learns a higher-order decision boundary,
$\hat{y} = \arg\max_{k \in \mathcal{C}} g_k(p,\ p - t)$. It recovers most of what the tuned
rules find, but it is a second model to train, calibrate, and maintain, which matters when
the rarest classes have only a few dozen validation examples, as here.

\begin{table}[t]
\caption{Decision engines on the test split (\M{3}); thresholds tuned on validation (\M{1}).}
\label{tab:policies}
\footnotesize
\setlength{\tabcolsep}{3pt}
\begin{tabular}{@{}lccccc@{}}
\toprule
\textbf{Decision rule} & \textbf{Acc.} & \textbf{Mac.-F1} & \textbf{Mac.-rec.} & \textbf{Routing} & \textbf{$\Delta$Mac.-F1} \\
\midrule
argmax (baseline)          & 0.983 & 0.295 & 0.247 & 0.983 & --- \\
absolute lift              & 0.984 & 0.324 & 0.267 & 0.984 & $+0.029$ \\
relative lift              & 0.987 & 0.422 & 0.407 & 0.989 & $+0.127$ \\
benign floor $\to$ argmax  & 0.984 & 0.358 & 0.316 & 0.985 & $+0.063$ \\
benign floor $\to$ abs.\ lift & 0.984 & 0.321 & 0.271 & 0.984 & $+0.026$ \\
benign floor $\to$ rel.\ lift & \textbf{0.992} & \textbf{0.489} & \textbf{0.520} & \textbf{0.996} & $+0.194$ \\
meta-stacker               & 0.988 & 0.435 & 0.445 & 0.991 & $+0.140$ \\
\bottomrule
\end{tabular}
\end{table}

Table~\ref{tab:policies} summarises the results. The strongest policy, \textbf{\#5, benign
floor $\to$ relative lift}, raises macro-F1 from $0.295$ to $0.489$. Improving recall on
rare classes would ordinarily be expected to cost precision on the majority class, yet
accuracy and routing accuracy also improve. This indicates that the naive argmax was
misclassifying a substantial number of rare attacks \emph{as} benign, consistent with the
\emph{positive} $\Delta$macro-F1 of every policy.

\begin{table*}[t]
\caption{Per-class performance on the test split (\M{3}): one-vs-rest PR-AUC (ranking) and
F1 under each decision engine. Grouped by whether ranking and decision agree. Best F1 per
class in bold.}
\label{tab:perclass}
\small
\begin{tabular}{@{}l c ccccccc@{}}
\toprule
 & & \multicolumn{7}{c}{\textbf{F1 by decision engine}} \\
\cmidrule(l){3-9}
\textbf{Class} & \textbf{PR-AUC} & argmax & abs.\ lift & rel.\ lift & floor$\to$argmax & floor$\to$abs. & floor$\to$rel. & meta \\
\midrule
\multicolumn{9}{@{}l}{\emph{Good ranking, compatible with argmax}} \\
\cls{benign}            & 1.000 & 0.992 & 0.992 & 0.995 & 0.993 & 0.992 & \textbf{1.000} & 0.996 \\
\cls{dos}               & 0.661 & \textbf{0.616} & 0.592 & 0.561 & 0.571 & 0.585 & 0.513 & 0.381 \\
\cls{bot\_farm}         & 0.801 & 0.475 & 0.641 & 0.444 & \textbf{0.695} & 0.667 & 0.379 & 0.634 \\
\cls{credential\_abuse} & 0.615 & 0.451 & 0.370 & 0.289 & \textbf{0.608} & 0.426 & 0.337 & 0.509 \\
\midrule
\multicolumn{9}{@{}l}{\emph{Good ranking, incompatible with argmax}} \\
\cls{jailbreak}         & 0.883 & 0.006 & 0.119 & 0.826 & 0.140 & 0.071 & \textbf{0.830} & 0.700 \\
\cls{distillation}      & 0.794 & 0.000 & 0.005 & 0.018 & 0.022 & 0.009 & \textbf{0.653} & 0.271 \\
\midrule
\multicolumn{9}{@{}l}{\emph{Weak ranking}} \\
\cls{harmful\_use}      & 0.249 & 0.040 & 0.100 & 0.400 & 0.095 & 0.071 & \textbf{0.466} & 0.332 \\
\cls{data\_extraction}  & 0.135 & 0.073 & 0.085 & \textbf{0.244} & 0.093 & 0.063 & 0.184 & 0.080 \\
\cls{agentic\_misuse}   & 0.006 & 0.000 & 0.010 & 0.022 & 0.008 & 0.002 & \textbf{0.042} & 0.009 \\
\bottomrule
\end{tabular}
\end{table*}

Per class (Table~\ref{tab:perclass}), key attack vectors improve substantially:
\cls{distillation} F1 rises from $0.00$ to $0.65$ and \cls{jailbreak} from $0.006$ to
$0.83$. The per-class results also reveal trade-offs:
\begin{enumerate}
  \item The best overall policy, \textbf{\#5}, lowers \cls{dos} F1 from $0.616$ to
    $0.513$, and \cls{credential\_abuse} and \cls{bot\_farm} performance also declines.
  \item \textbf{\#3, benign floor $\to$ argmax}, shows the opposite trade-off. It improves
    the middle group (\cls{bot\_farm} $0.475 \to 0.695$, \cls{credential\_abuse}
    $0.451 \to 0.608$) but barely changes the well-ranked yet rarely selected classes such
    as \cls{distillation}.
  \item No policy substantially improves \cls{agentic\_misuse} or \cls{data\_extraction},
    whose best F1 scores are $0.042$ and $0.244$ respectively.
  \item \cls{harmful\_use} is an exception: despite a middling ranking (PR-AUC $0.25$), it
    reaches an F1 of $0.47$ under the best policy.
  \item \textbf{\#6, the meta-stacker}, performs well, with a macro-F1 of $0.44$ and the
    most consistent performance across the nine classes. It may be preferable where labels
    are more abundant and the added complexity of a meta-model is acceptable.
\end{enumerate}
No single policy is best for every class; each represents a different trade-off between
the well-separated classes and the difficult ones.

\subsection{Feature Importance}

\begin{figure}[t]
\centering
\begin{tikzpicture}
\begin{axis}[
  xbar, width=0.5\columnwidth, height=7.6cm, scale only axis=false,
  xmin=0, xmax=0.7, y dir=reverse, ytick={1,...,15},
  yticklabels={
    d7\_rolling\_mean\_n\_requests,
    domain\_coverage\_cumulative,
    noisy\_registration\_velocity,
    is\_api,
    d7\_rolling\_max\_noisy\_max\_safety\_score,
    avg\_inter\_request\_time,
    d7\_rolling\_mean\_noisy\_std\_inter\_request\_time,
    d7\_rolling\_n\_captcha,
    noisy\_shared\_device\_count,
    noisy\_shared\_ip\_count,
    is\_datacenter\_ip,
    d7\_rolling\_mean\_total\_tokens,
    d7\_rolling\_session\_count,
    geo\_region,
    d7\_rolling\_mean\_noisy\_avg\_safety\_score},
  yticklabel style={font=\ttfamily\tiny},
  xticklabel style={font=\tiny, /pgf/number format/fixed},
  xtick={0,0.2,0.4,0.6},
  xlabel={mean $|\mathrm{SHAP}|$}, xlabel style={font=\scriptsize},
  bar width=5pt, enlarge y limits=0.04,
  axis x line*=bottom, axis y line*=left,
  nodes near coords, nodes near coords style={font=\tiny, text=black,
    /pgf/number format/fixed, /pgf/number format/precision=3},
]
\addplot[fill=black!35, draw=black!70] coordinates {
  (0.539,1) (0.392,2) (0.268,3) (0.255,4) (0.152,5) (0.150,6) (0.115,7) (0.115,8)
  (0.097,9) (0.092,10) (0.087,11) (0.076,12) (0.076,13) (0.069,14) (0.052,15)};
\end{axis}
\end{tikzpicture}
\caption{Global feature importance for the attack-type GBDT: mean $|\mathrm{SHAP}|$ over a
200k-row test sample.}
\label{fig:shap}
\Description{Horizontal bar chart of mean absolute SHAP values for the fifteen most important features, led by the seven-day rolling mean request count.}
\end{figure}
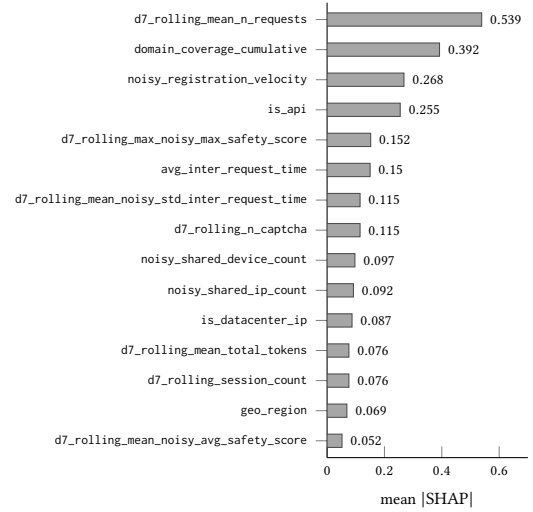

To check which features the model relies on, and whether its behaviour is plausible, we
compute SHAP values~\cite{lundberg2017shap} (Figure~\ref{fig:shap}). The most important
feature is request \emph{volume} (the seven-day mean request count), followed by cumulative
\emph{domain coverage}, a natural signal for distillation rings that systematically
harvest the model's knowledge, consistent with the mechanism encoded in the SCM. Next is a
cluster of \emph{infrastructure-graph} signals (registration velocity, shared devices and
IPs, and datacenter and API flags), characteristic indicators of coordinated fake-account
fleets. Timing cadence and peak safety scores complete the top of the list. The model thus
recovers heuristics and priors that were built into the SCM; the extent to which this
reflects circularity between generator and detector is a question we leave open.

% ============================================================== CLOSING ==
\section{Conclusion}
\label{sec:closing}

We studied inference-layer defence for a hypothetical LLM provider, \fiveel, starting at
the second rung of the Ladder of Abstraction: a per-session point predictor, trained on a
synthetic dataset from a structural causal model that simulates coordinated campaigns,
platform feedback, and three tiers of label observability. Two findings stand out. First,
the labels used for training and evaluation matter as much as the model: the same detector
is near-perfect against oracle labels and mediocre against operational ones, and the gap
directly measures the value of a labelling pipeline. Second, a model that ranks rare
attacks well can still fail to \emph{select} them under an overwhelming benign prior; a
simple thresholded decision engine recovers much of this lost signal.

\paragraph{Limitations.} Our results are obtained on synthetic data whose generator fixes
account roles, models label delay as a per-type prior rather than an account-level
mechanism, and encodes the same heuristics that the detector later recovers; real traffic
is likely to be harder. The weighting and decision rules are heuristic and tuned on a
single dataset. Finally, the remaining failures, \cls{agentic\_misuse} and
\cls{data\_extraction}, call for more data or for the higher rungs of the ladder, which we
leave to future work.

\section*{Data Availability}
The full dataset, including the raw user-session log with all three label channels and
the engineered feature matrix, is publicly available on Hugging
Face~\cite{daoistdurian2026dataset} at
\url{https://huggingface.co/datasets/DaoistDurian/lang-adversarial-inference-01}.

% =========================================================== REFERENCES ==
\bibliographystyle{ACM-Reference-Format}
\bibliography{references}

% ============================================================= APPENDIX ==
\appendix

\section{Dataset Sample}
\label{app:sample}

Table~\ref{tab:sample} shows one raw user-session record per attack type, with a subset of
columns. \cls{label} is the \M{2} channel; ``---'' marks a session that is unlabelled on
that channel (for \cls{benign} accounts, the attack type itself is empty). The full dataset,
including the engineered feature matrix with the materialised \M{1}/\M{2}/\M{3} label
triplets and the trailing seven-day rolling features, is available on Hugging
Face~\cite{daoistdurian2026dataset}.

\begin{table*}[h]
\caption{A stratified sample of raw user-session records (oracle \M{3} fields included).
IRT = inter-request time (s); QSE = query-structure entropy; cov.\,$\Delta$ = domain
coverage delta.}
\label{tab:sample}
\footnotesize
\setlength{\tabcolsep}{4.5pt}
\begin{tabular}{@{}l l r r c r r r r r r l@{}}
\toprule
\textbf{Attack type} & \textbf{Phase} & \textbf{Day} & \textbf{Req.} & \textbf{API} &
\textbf{Avg IRT} & \textbf{Std IRT} & \textbf{QSE} & \textbf{Safety} & \textbf{Cov.\,$\Delta$} &
\textbf{Soph.} & \textbf{Label (\M{2})} \\
\midrule
\cls{benign}            & dormant      & 9   & 6   & --         & 22.74 & 14.94 & 0.214 & 0.001 & 0.000 & 0.00 & --- \\
\cls{bot\_farm}         & active       & 104 & 17  & \checkmark & 2.15  & 0.67  & 0.069 & 0.005 & 0.000 & 0.12 & \cls{bot\_farm} \\
\cls{dos}               & active       & 1   & 382 & \checkmark & 0.49  & 0.43  & 0.014 & 0.035 & 0.039 & 0.60 & \cls{dos} \\
\cls{credential\_abuse} & exploitation & 36  & 26  & \checkmark & 8.37  & 2.92  & 0.747 & 0.079 & 0.000 & 0.65 & \cls{credential\_abuse} \\
\cls{distillation}      & active       & 27  & 28  & \checkmark & 3.84  & 15.53 & 0.069 & 0.002 & 0.000 & 0.40 & --- \\
\cls{jailbreak}         & cooldown     & 59  & 5   & --         & 30.10 & 11.30 & 0.693 & 0.028 & 0.000 & 0.84 & \cls{jailbreak} \\
\cls{harmful\_use}      & active       & 22  & 9   & --         & 3.60  & 10.95 & 0.067 & 0.478 & 0.000 & 0.34 & --- \\
\cls{data\_extraction}  & active       & 19  & 23  & \checkmark & 12.69 & 2.99  & 0.012 & 0.106 & 0.000 & 0.88 & --- \\
\cls{agentic\_misuse}   & probing      & 57  & 18  & \checkmark & 3.73  & 0.62  & 0.382 & 0.040 & 0.000 & 0.70 & --- \\
\cls{cover\_traffic}    & probing      & 93  & 7   & --         & 50.06 & 24.29 & 0.393 & 0.002 & 0.000 & 0.70 & --- \\
\bottomrule
\end{tabular}
\end{table*}

\end{document}